\documentclass[11pt]{article}
\usepackage{xcolor}
\usepackage{natbib}
\usepackage{amssymb}
\usepackage{amsmath}
\usepackage{graphicx}
\usepackage{booktabs}
\usepackage{threeparttablex}
\usepackage{lscape}
\usepackage{authblk}

\definecolor{Black}{RGB}{0,0,0}
\title{Predictive Effects of Novelty Measured by Temporal Embeddings on the Growth of Scientific Literature} 
\author[1]{Jiangen He and Chaomei Chen \thanks{jiangen.he@drexel.edu, chaomei.chen@drexel.edu. Department of Information Science, Drexel University.}}
\begin{document}

\maketitle

\begin{abstract} 
Novel scientific knowledge is constantly produced by the scientific community. Understanding the level of novelty characterized by scientific literature is key for modeling scientific dynamics and analyzing the growth mechanisms of scientific knowledge. Metrics derived from bibliometrics and citation analysis were effectively used to characterize the novelty in scientific development. However, time is required before we can observe links between documents such as citation links or patterns derived from the links, which makes these techniques more effective for retrospective analysis than predictive analysis. In this study, we present a new approach to measuring the novelty of a research topic in a scientific community over a specific period by tracking semantic changes of the terms and characterizing the research topic in their usage context. The semantic changes are derived from the text data of scientific literature by temporal embedding learning techniques. We validated the effects of the proposed novelty metric on predicting the future growth of scientific publications and investigated the relations between novelty and growth by panel data analysis applied in a large-scale publication dataset (MEDLINE/PubMed). Key findings based on the statistical investigation indicate that the novelty metric has significant predictive effects on the growth of scientific literature and the predictive effects may last for more than ten years. We demonstrated the effectiveness and practical implications of the novelty metric in three case studies.
\end{abstract}

\section{Introduction}

Novelty and growth are two widely used attributes for characterizing how a research topic emerges in science \citep{tu2012indices, Small2014, Rotolo2015}. Novelty and growth are likely to co-evolve but evolve along almost inverse paths over different stages of the emergence of a research topic. \citet{Rotolo2015} qualitatively and \citet{tu2012indices} quantitatively depicted the co-evolution of the two attributes. At the stage right before its emergence, a research topic is characterized by a high level of novelty, but doesn't attract much attention from the scientific community and its growth is relatively low as a consequence of the limited impact. After the appearance of certain turning points (critical scientific publications that re-directs the flow of science) \citep{Chen2004}, the research topic starts to take off and grows fast, but the level of novelty will decrease gradually once the emergence becomes apparent. After acquiring a rapid growth at the stage of emergence, the scientific knowledge of the research topic become well-established, and the level of novelty is likely to decrease even further at the post-emergence stage. According to this basic model, novelty may be an early sign of growth and emergence of a research topic. Many existing studies identify or predict emerging topics by characterizing novelty from various dimensions. However, few studies quantitatively analyzed how novelty affects growth in science.

Measuring novelty manifested by scientific literature is the first challenge for understanding the effects of novelty on the growth of science. Citation analyses were commonly and effectively used for identifying novelty \citep{small2006tracking, shibata2009comparative,   glanzel2012using, Small2014}. The basic idea of these studies is that clusters of documents or words represent different scientific areas and new clusters or new content of clusters characterize the novelty of the scientific areas. A relatively coherent set of publications related to a certain research topic is necessary for these techniques to detect the cluster and novelty of the research topic, which requires time to attract researchers to devote to this topic and produce related publications. Therefore, these techniques are intrinsically insensitive to the novelty in scientific literature because of the time lag. For example, \cite{Small2014} effectively nominated a list of emerging topics by building citation networks based on a large-scale publication dataset, but many of identified topics were well recognized by the scientific community before the time of emergence identified by their approaches. Compared to citation analyses, text analyses may reduce the time lag by directly measuring the novelty expressed by text data of scientific literature. 

The key challenge to derive novelty measure from the textual information of scientific literature is how to effectively and efficiently represent the semantics and the semantic changes of research topics without information loss. { \color{Black}Word embedding techniques, such as word2vec \citep{mikolov2013distributed} and GloVe \citep{pennington2014glove} have proved their utility in representing the semantics of words {\color{Black}, and techniques for learning semantic changes were also developed \citep{jurgens2009event, hamilton_diachronic_2016}}. \cite{hamilton_diachronic_2016} developed a temporal word embedding method to understand how the semantics of words changed over time, by aligning word embeddings across different periods. Based on the temporal word embedding method, we quantified the temporal semantic changes of research topics and used it as a proxy to measure the novelty of the research topics.} {\color{Black}In this study, the novelty of a research topic is defined as a measure of how much new scientific knowledge was produced by the scientific community and characterized by scientific literature on the research topic in a specific period. The novelty metric is operationalized into a metric of semantic changes of the term(s) describing the research topic in scientific literature.} Unlike previous citation-based methods that derive novelty measure from the dynamics of citation links, we quantify novelty by making use of rich textual information of scientific literature without relying on citation information. Thus, our approach is applicable to a wider range of data sources than approaches that solely rely on citation data. Furthermore, it is conceivable that an integrative approach may improve the effectiveness even further. Additionally, vectorizing and quantifying research concepts and their novelty can improve their interpretability because of the modeling of conceptual relationships. It also significantly widens their applicability with artificial intelligence techniques.

We also address an issue concerning whether novelty can serve as an early sign of future scientific growth and how it predicts growth. The growth of a certain research topic is defined as the growth rate of knowledge outputs on the topic, which is operationalized into the growth rate of publications on the topic in this study. {\color{Black}The research topics in this study are operationalized into research concepts selected from descriptors in a comprehensive controlled vocabulary of life sciences, Medical Subject Headings (MeSH).} We conducted an investigative study in a large collection of scientific publications (MEDLINE/PubMed) spanning 35 years. The investigated data span across topics and years. Panel data models can examine cross-sectional (research topic) as well as time-series (time) effects. Therefore, we conducted our regression analysis of panel data models. Based on the results of regression analysis, we validated the predictive effects of the measured novelty on the growth in science and statistically investigated how novelty affects the growth. 

We summarized our main contributions as below:
\begin{enumerate}
    \item We provided a new method for measuring the novelty of research topics through temporal embeddings.
    \item We validated the predictive effects of the proposed novelty metric on the growth of research topics.
    \item We statistically investigated how novelty effects the growth of research topics.
\end{enumerate}

\section{Related Work}

Direct citation analysis \citep{garfield1964use}, co-citation analysis \citep{small1973}, and bibliographic coupling \citep{kessler1963} were commonly used for identifying novelty in science development. These citation analysis techniques built based on citation patterns among the different links of scientific publications. \cite{erdi2013prediction} use the appearance of new patent co-clusters to represent new technology areas. The newly appeared areas naturally are more likely to be highly novel but existing areas also have the possibility to gain high novelty in the evolution of science. For this reason, many studies used various metrics to characterize the novelty of citation clusters. For examples, \cite{small2006tracking} defined a simple metric named \textit{currency} to characterize the \textit{newness} of a co-citation cluster and found the \textit{currency} variable had predictive effects on the growth of clusters; \cite{shibata2011detecting} used topological measures to determine whether there are novel clusters of citation network and analyzed the clusters with the average published and parent-child relationship to detect their trends. Different types of citation differ in mapping research fronts and reflect different organizing principles \citep{boyack2010co,shibata2009comparative} so hybrid methods aimed to make use of different characteristics of types of citation to detect novelty. \cite{glanzel2012using} introduced three paradigmatic types of new topics by combining three types of citation network. \cite{Small2014} detected growth and novelty based on a combination of direct citation and co-citation networks where the size changes of direct citation cluster reflect the growth rates, and the number of papers of direct clusters, which are also in new co-citation threads, reflects the novelty. Citation analysis is efficient and has the potential to detect research topics automatically, but the formation of citation clusters requires time, and the results of citation analysis are less semantically interpretable.

Many studies used text mining techniques that scan a large volume of textual data to identify the degree of novelty in science and technology. \cite{lee2015novelty} used text mining to extract the patterns of word usage and adopted local outer factor (LOF) to measure novelty of patents. Based on the identified novel patents and patent mapping, technologies with novelty and opportunities may be explored and identified. More studies focus on topic-level analysis based on topic modeling techniques \citep{blei2003latent,blei2007correlated}. \cite{morchen2008anticipating} tracked the frequency of topics over time and used the frequency score to indicate novelty of topics. Some studies \citep{he2009detecting,yan2014research} employed a topic's temporal relationship with other topics to decide the newness of topics{\color{Black}, but their methods of building the relationship of topics are different}. \cite{he2009detecting} adapted topic modeling to citation networks to specify the pairwise relationship, but \cite{yan2014research} used a similarity measurement to build the relationship. These techniques can automatically detect research topics based on textual information and identify their novelty, but few studies investigated the impact of the novelty degree of research topics on the growth of scientific knowledge.

Another study \citep{tu2012indices}, which also explicitly discussed scientific growth and the degree of novelty as ours, utilized neither citation links nor textual data. Their study defined a novelty index and a publish volume index which are utilized to determine the detection points of new emerging topics.

\section{Measuring Novelty of Research Topics} \label{section:measure_novelty}

In this section, we outline how we train temporal word embeddings models on processed MEDLINE/PubMed data, by first constructing embeddings model in each period and then aligning them over time. We also proposed a metric that we used to quantify the novelty degree of a research topic.

\subsection{Data and preprocessing}
The text dataset we used to train our model is taken from MEDLINE/PubMed. The MEDLINE/PubMed data contains over 26 million journal citations and abstracts for biomedical literature from around the world which is often cited as the largest database of biomedical publications \citep{USNationalLibraryofMedicine}. We used the baseline set of MEDLINE/PubMed released in December 2016 for training word embedding models. {\color{Black}The titles and abstracts of the biomedical articles were extracted for training our model. The abstracts of 36.36\% articles are not provided by MEDLINE/PubMed, but almost all of them have title information.}

Besides the large volume of the dataset, another desirable feature of it for our study is the process of indexing. {\color{Black}Most of publications in this database (88.25\%\footnote{https://www.nlm.nih.gov/bsd/licensee/2017\_stats/2017\_LO.html}) are} indexed by a set of descriptors from MeSH which can be used to improve the model training and facilitate the evaluation of our experiment.

We conduct a preprocessing by using a Python library \textit{NLPre} \footnote{https://pypi.python.org/pypi/nlpre} as below
\begin{enumerate}
    \item Conduct a series of preprocessing steps to remove noise and errors including dash removal, URL replacement, capitalization normalization, etc. 
    \item Replace phrases from MeSH dictionary. MeSH provides a list of `Entry Terms' for each MeSH descriptor, which are synonyms, alternate forms, and other closely related terms of the MeSH descriptor. {\color{Black}Biomedical concepts will be replaced by a unified single-word term. For example, `AIDS Antibodies', `HTLV III Antibodies', `HIV Associated Antibodies', and other 9 synonyms of `HIV' will be replaced by `HIV\_Antibodies'\footnote{https://meshb.nlm.nih.gov/record/ui?ui=D015483}. This step is crucial not only for resolving the issue of synonyms but also for tracking the semantic changes of research topics since we cannot obtain a representation of a multi-word term by original word2vec.}
    \item Identify acronyms of phrases found in a parenthesis and replace all instances of acronyms with the given phrases. For example, through the text snippet of `Environmental Protection Agency (EPA)' in a document, EPA will be identified as the acronym of `Environmental Protection Agency' and all acronyms of `EPA' will be replaced by `Environmentz\_Protection\_Agency' in this document.
\end{enumerate}

\subsection{Temporal embeddings} 
We need to identify and measure new information expressed by scientific literature to quantify the novelty of research topics. Since word embedding techniques can be effective and efficient in capturing syntactic and semantic relationships, we adopt semantic features derived from word embedding models to identify and measure the new information on research topics in a specific period. We trained a word embedding model for each period and then align embedding models sequentially. Here we elaborate on how we construct word2vec models \citep{mikolov2013distributed} trained on MEDLINE/PubMed data and quantify novelty of research topics based on these trained models.

\subsubsection{Embedding learning}
Our goal in this step is to get the contextual embedding of research concepts from publication data. We used skip-gram with negative sampling (SGNS) introduced by \cite{mikolov2013distributed} to learn research concept embedding based on the context words of research concepts. Given a word or phrase \(w\) in training dataset, skip-gram maps it into a continuous representation $\mathbf{w}$. $\mathbf{w}$ is used to predict the context words of $w$. {\color{Black}The objective of skip-gram is to maximize the log probability:{\color{Black}
\begin{equation}
	\sum_{(c, w) \in D } \log p(D=1|c, w)
\end{equation}
where $c$ is the context of $w$ and $D$ is the set of all of pairs in the training data. Instead of looping over the entire words in training, negative sampling can be used to generate the set of $D'$ of random $(c, w)$ pairs which are not in the training data to accelerate the training procedure. Through negative sampling, new objective would be to maximize the log probability:
\begin{equation}
	\sum_{(c, w) \in D } \log p(D=1| c, w) + \sum_{(c, w) \in D' } \log p(D=0| c, w)
\end{equation}
}} 

We separately constructed embeddings of research concepts from publication text data for each period by SGNS algorithm. {\color{Black}We used the implementation of word2vec provided by gensim \citep{rehureklrec} for embedding learning. We empirically set embedding length as 100, window size as 5, negative sampling size as 5, and the number of iteration as 5}.

\subsubsection{Alignment of embeddings}
The embeddings constructed in different time periods are in different vector space {\color{Black}because of differences in stochastic initialization of the weights of the neural network in SGNS algorithm}. The different vector spaces precluded the comparison of the same research concepts across periods. In order to compare vectors from different periods, embeddings from different time periods need to be aligned into the same coordinate axes. We use orthogonal Procrustes to align the learned low-dimensional embeddings as \cite{hamilton_diachronic_2016}. Defining \( \mathbf{W}^{(t)} \in \mathbb{R}^{d \times \left| \nu \right|} \) as the matrix of word embeddings learn at period \(t\), we align across time periods while preserving cosine similarities by optimizing
\begin{equation}
	\mathbf{R}^{(t)} = \textup{arg} \underset{\mathbf{Q}^\textup{T}\mathbf{Q}=\mathbf{I}}{\textup{min}}\left \| \mathbf{Q}\mathbf{W}^{(t)} - \mathbf{W}^{(t+1)} \right \|_F
\end{equation}
with \(\mathbf{R}^{(t)} \in \mathbb{R}^{d \times d}\). {\color{Black} The alignment is performed in an iterative fashion, i.e., ($\mathbf{W}^{(1)}$, $\mathbf{W}^{(2)}$), ($\mathbf{W'}^{(2)}$, $\mathbf{W}^{(3)}$), ..., ($\mathbf{W'}^{(T-1)}$,$\mathbf{W}^{(T)}$) where $\mathbf{W'}^{(t)}$ is the aligned matrix of word embeddings at $t$, an alignment of ($\mathbf{W'}^{(t-1)}$, $\mathbf{W}^{(t)}$) produces an aligned matrix $\mathbf{W'}^{(t)}$, and $T$ is the last time-period. }

\subsection{Novelty} \label{section:novelty}

After aligning the embeddings for each period, we employ the aligned embeddings to measure the novelty of a research topic in a specific time period. Since the embeddings were aligned, the vectors representing words or phrases in different time periods are in the same vector space, which enables linear algebraic computation between them. The negative of the cosine of the angles between a research topic $i$ vector $\mathbf{w}_i^{(t)}$ at period $t$ and each previous vectors $\mathbf{w}_i^{(t-\Delta t, ..., t-1)}$ determine the novelty score for that topic $i$ at $t$ \citep{allan2003retrieval}.
We compute the cosine similarity (\textup{cos-sim}) to make the comparison of semantic properties and then measure novelty of a research topic \(w_i\) at time period \(t\) as follows:
\begin{equation}
	Novelty^{(t)}_i(win) = 1 - \underset{0 < \Delta t < win}{\max}\textup{cos-sim}(\mathbf{w}_i^{(t-\Delta t)}, \mathbf{w}_i^{(t)})
\end{equation}
with different windows \( win \) (i.e., how many periods in the past to compare with).

\section{Case Studies}
Before statistically investigating the relationship between novelty and growth in science, {\color{Black}we show three examples to initially verify our methods. We chose two notable viruses which had lead to two large-scale disease breakouts recently, namely \textit{Ebola Virus} and \textit{H1N1 Virus} as examples. Additionally, we chose \textit{Peptic Ulcer} as the third example because it was associated with Nobel Prize in 2005. Their MeSH descriptors are \textit{Ebolavirus} (D029043)\footnote{https://meshb.nlm.nih.gov/record/ui?ui=D029043}, \textit{Influenza A Virus, H1N1 Subtype} (D053118)\footnote{https://meshb.nlm.nih.gov/record/ui?ui=D053118}, and \textit{Peptic Ulcer} (D010437)\footnote{https://meshb.nlm.nih.gov/record/ui?ui=D010437}. 

In each case study, we used our methods to learn the novelty score changes of a research topic in a specific observed period and analyzed temporal patterns in the co-evolution of novelty and growth of the research topic. {\color{Black}The novelty score is the value of $Novelty(7)$, i.e., our proposed novelty metric with a seven-year window.} To learn novel ideas behind the novelty score changes of a specific research topic, we show a temporal visualization of scientific evolution of the research topic. We used a modified visualization method based on the one introduced by \cite{hamilton_diachronic_2016}. Firstly, we selected most related terms of the research topic for each year within the observed period and used these terms to visualize the research development of each year. However, some terms may repeatedly be related to the research topic for different years. To reduce reductant information, we only retained the repeated related terms with changed semantic meaning. Specifically, for a related term $w$ at $t$ once appeared at $t'$ where $t' < t$, the similarity between $w^{(t)}$ and $w^{(t')}$ determines if $w$ would be retained at $t$. Only when $\textup{cos-sim}(w^{(t)},w^{(t')}) < 0.5$, $w$ would be retained at $t$. Then, we computed t-SNE \citep{maaten2008visualizing} two-dimensional embedding of terms over each year to mapping the terms into visual space. Lastly, we encoded the color of terms by the year they were selected and used the force-based collision detection to remove the overlaps between terms. However, fully interpreting the visualization may be beyond our domain expertise and domain experts may obtain more insights from the visualization.}

\subsection{Case I: Ebola Virus}
{\color{Black}Ebola virus disease is a severe, often fatal illness in human. It first appeared in 1979 in 2 simultaneous outbreaks in African countries. The 2014-2016 outbreak in West Africa was the largest and most complex outbreak since its discovery \citep{who:elbola}. We chose research on \textit{Ebola Virus} from 2011 to 2015 as the example topic.}

\begin{figure}[!ht]
  \centering
  \includegraphics[width=0.8\linewidth]{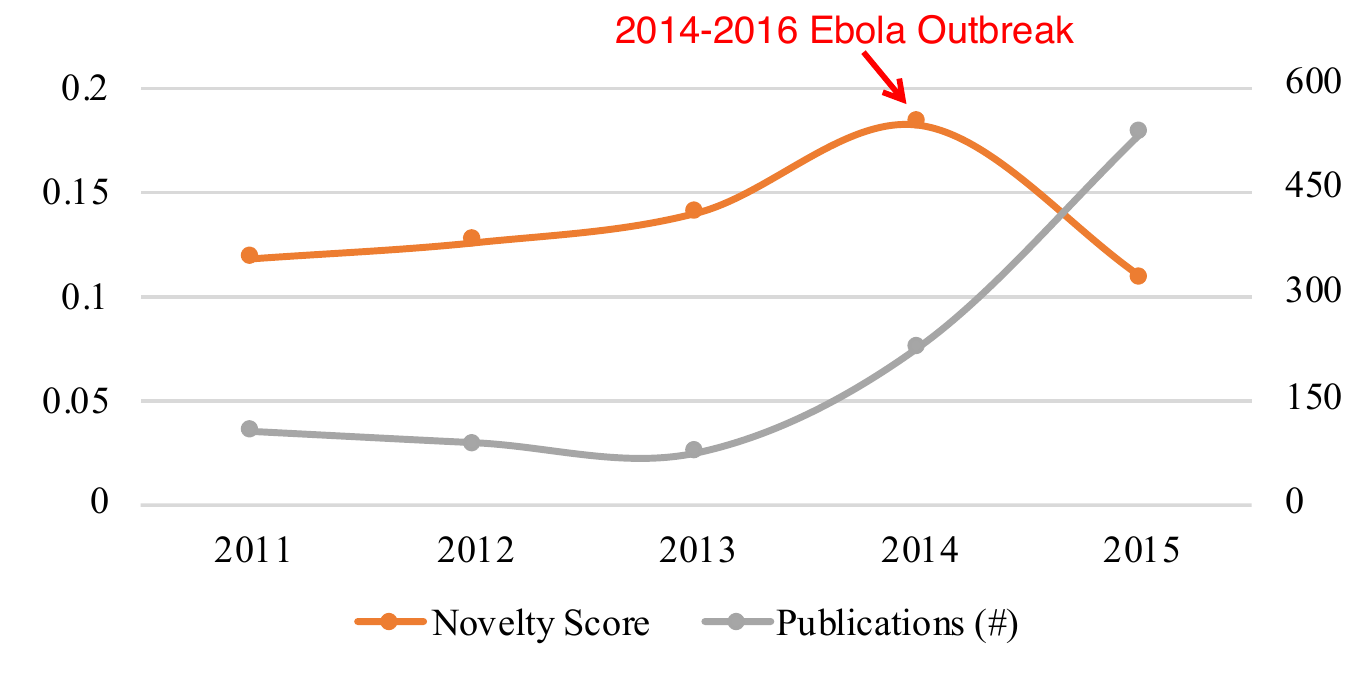}
  \caption{\label{fig:ebola_curve} Co-evolution of the novelty and growth of research on `Ebola Virus'.}
\end{figure}

Figure \ref{fig:ebola_curve} shows the novelty and growth co-evolution of research on \textit{Ebolavirus} from 2011 to 2015. From Figure \ref{fig:ebola_curve}, we can see the relatively high novelty of research on \textit{Ebolavirus} in 2013 and 2014 was followed by the rapid growth in 2014 and 2015. Following the growth in 2014 and 2015, the novelty decreased in 2015. The co-evolving pattern of growth and novelty is in line with the pre-emergence and emergence stages described at the beginning of this article.

\begin{figure}[!ht]
  \centering
  \includegraphics[width=1\linewidth]{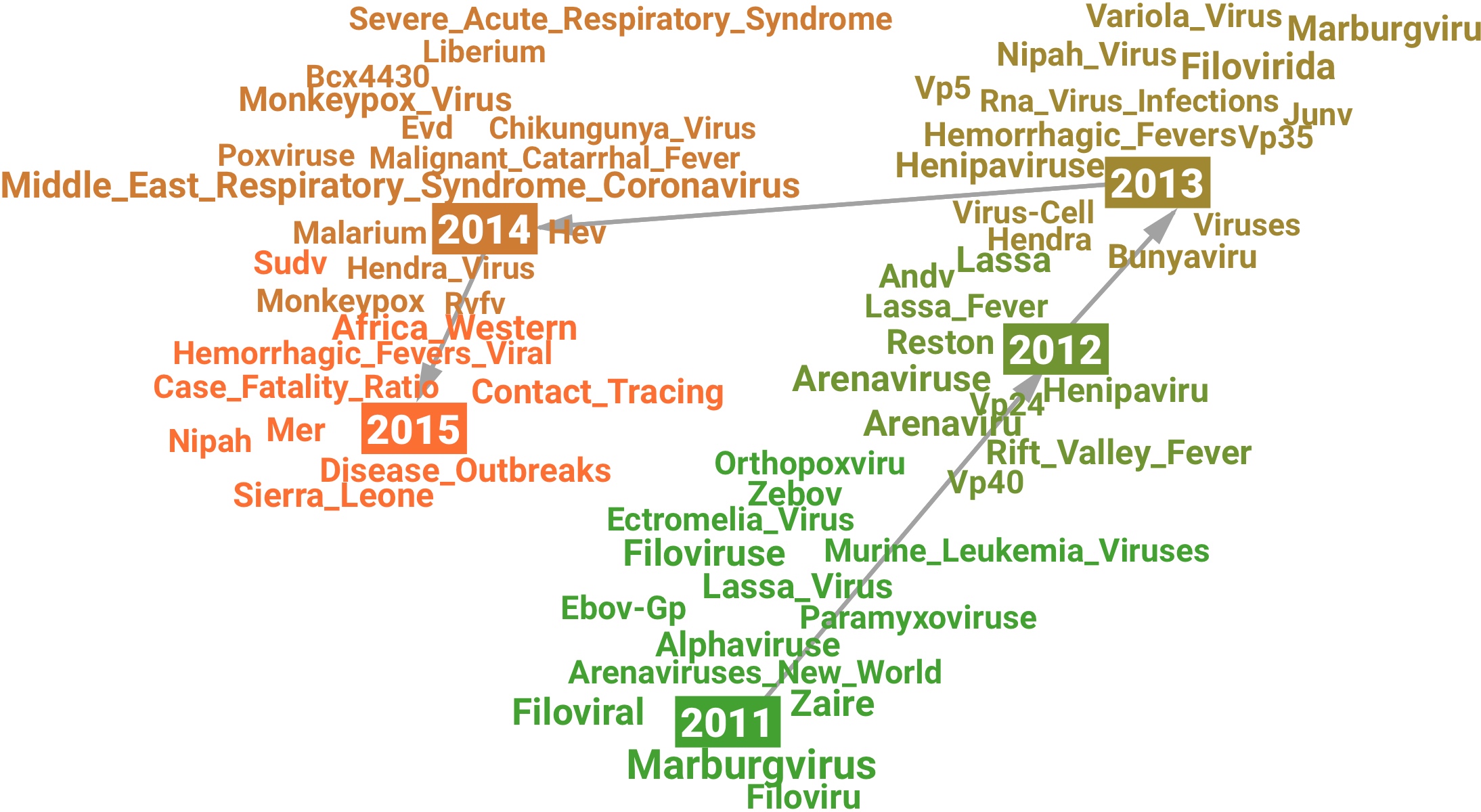}
  \caption{\label{fig:ebola_vis} Visualization of the semantic change in research on `Ebola Virus' from 2011 (green) to 2015 (orange).}
\end{figure}

The visualization (see Figure \ref{fig:ebola_vis}) shows a picture of how research on \textit{Ebola Virus} developed from 2011 to 2015. A significant change may have happened in research on \textit{Ebola Virus} in 2014 which is indicated by a new cluster formed after 2014 in the visualization as well as the high novelty score in 2014 (see Figure \ref{fig:ebola_curve}). The related terms for each year may indicate the changing research content over time. For example, {\color{Black}the term \textit{Variola\_Virus} emerged in 2013 because its context was similar as \textit{Ebola Virus}'s in scientific literature.} The reason behind the emergence may be that the science community `expected to eradicate Ebola virus by a safe and efficient vaccine development similar to the case of smallpox virus which was extinguished from the world by the variola vaccine' \citep{hong2014ebola}.

{ \color{Black}

\subsection{Case II: H1N1 Virus}
Influenza A (H1N1) virus is the subtype of influenza A virus that was the most common cause of \textit{2009 flu pandemic}. We chose the research on H1N1 from 2003 to 2014 as the example of this case study.

Figure \ref{fig:h1n1_curve} shows the changes of novelty and publication growth in research on the H1N1 virus from 2003 to 2014. The temporal patterns of novelty and growth are more complex than the case of Ebola Virus. We can see two peaks of novelty in 2004 and 2009 respectively. Both peaks of novelty were followed by the rapid growth of publications that occurred in the periods from 2006 to 2008 and from 2009 to 2010 respectively. However, the temporal patterns of these two pairs of novelty-growth relationships were different. The rapid growth from 2006 to 2008 was observed two years after the novelty peak in 2004, but the growth spurt in 2009 occurred in the same year as the novelty peak in 2009. The difference may be caused by the global breakout of the 2009 Flu Pandemic \citep{wiki:2009}. Due to this notable social factor, we observed both of growth and novelty spurt in the same year of 2009.
\begin{figure}[!ht]
  \centering
  \includegraphics[width=0.9\linewidth]{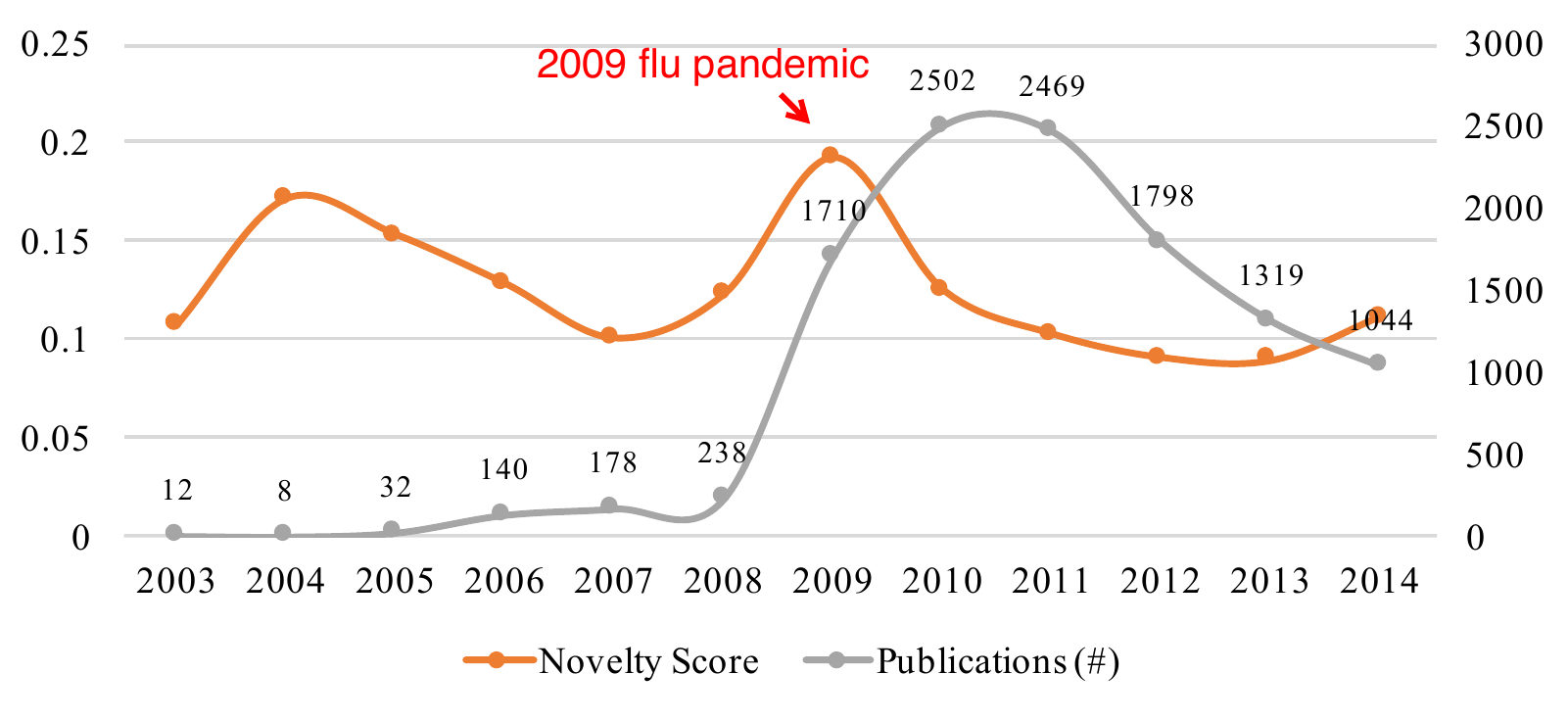}
  \caption{\label{fig:h1n1_curve} Co-evolution of the novelty and growth of research on H1N1 Virus.}
\end{figure}

The visualization in Figure \ref{fig:h1n1_vis} shows the evolution of research on H1N1 Virus from 2003 to 2014. We can observe three common features across these years: a) many other subtypes of Influenza virus A appeared in different years which might indicate related research interests in this area emerged, such as H6N2 in 2003, H9N2 in 2005, and H3N2 in 2009; b) several geographical terms were highlighted in different year because this research area is usually motivated by the breakouts of H1N1 virus in specific places, such as Guangdong in 2007 and Panama in 2008; c) the origination of virus was a constantly investigated research question according to the animal related terms in different years, such as Duck in 2007 and Swine-Origin in 2009. 

By analyzing the terms in 2004 and 2009, we may learn some ideas about the novelty. For example, Swine-Origin and S-Oiv in 2009 may indicate research on a new swine-origin influenza A (H1N1) virus (S-OIV) emerged in North America in early 2009 \citep{smith2009origins}. As another interesting example, we see the term of \textit{oseltamivir-resistant} both in 2004 and 2009, which means research activities on oseltamivir-resistant and H1N1 were active but in a different research context in these two years, according to our visualization method introduced at the beginning of this section. In 2004, although one patient with oseltamivir-resistant novel H1N1 was identified in Denmark, this didn't change the recommendations, made by the U.S. Centers for Disease Control and Prevention (CDC), on using oseltamivir on antiviral treatment of influenza A \citep{tucker2004acip}. However, oseltamivir resistance increased significantly for the first time worldwide during the 2007-2008 influenza season \citep{dharan2009infections}. This change of antiviral drug resistance patterns might have also greatly changed research on the H1N1 virus in 2009, so we see oseltamivir-resistant reemerged in a new context in 2009.

\begin{figure}[!ht]
  \centering
  \includegraphics[width=1\linewidth]{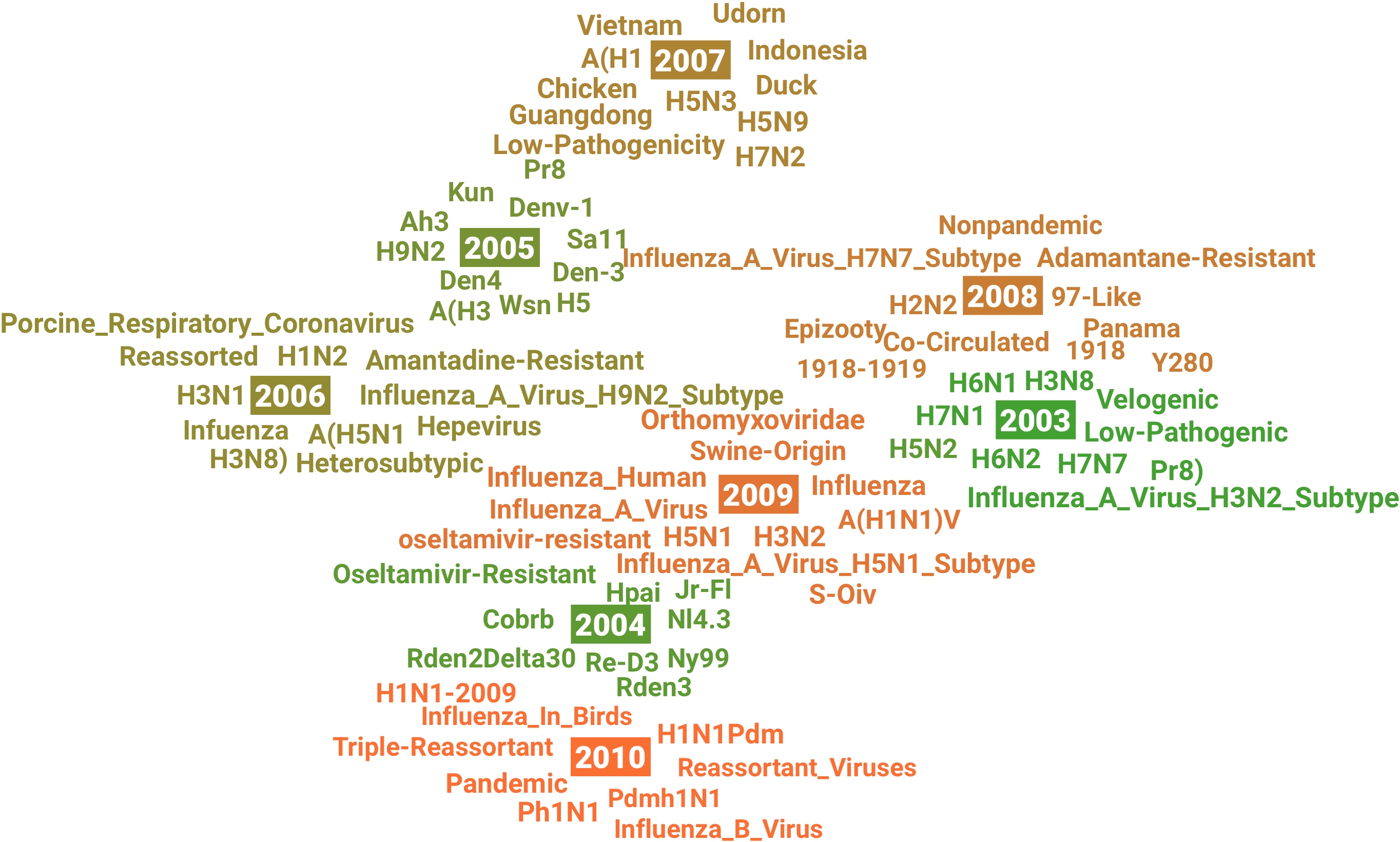}
  \caption{\label{fig:h1n1_vis} Visualization of the semantic change in research on `H1N1 Virus' from 2003 (green) to 2010 (orange).}
\end{figure}

\subsection{Case III: Peptic Ulcer} \label{section:case_peptic}
The Nobel Prize in Physiology or Medicine for 2005 was jointly awarded to Barry J. Marshall and J. Robin Warren for their paradigm-shifting discovery of `the bacterium Helicobacter pylori and its role in gastritis and peptic ulcer disease'. Before the link between \textit{Helicobacter pylori} infection and subsequent gastritis and peptic ulcer disease has been established, stress and lifestyle were considered the major causes of peptic ulcer disease in scientific community. We chose \textit{Peptic Ulcer} as an example to learn the role of novelty played in a scientific paradigm shift.

In Figure \ref{fig:peptic_curve} we also see two pairs of novelty and subsequent growth within the period from 1982 to 1988, but we mainly analyzes the second one which might be brought by the discovery of Helicobacter pylori. The first major publication of the Helicobacter pylori was the article \citep{MARSHALL19841311} published in \textit{Lancet} in 1984 \citep{pincock2005nobel}, {\color{Black}but the novelty score in 1984 was relatively low because attracting attention to this discovery from the scientific community took time. The first novelty score peak after the publication of Marshall-1984 occurred in 1986. By analyzing the citations of Marshall-1984 from 1984 to 1990 (see the top of Figure \ref{fig:peptic_curve}), we may see how the novelty score peak was related the to Marshall-1984.} According to our previous study \citep{CHEN2009191}, the first citation burst period of Marshall-1984 was between 1986 and 1988. The beginning of the burst period was in the same year of the novelty peak in 1986. Moreover, about 10\% research articles on Peptic Ulcer published in 1986 cited Marshall-1984. Considering the revolutionary content of Marshall-1984, it is reasonable to infer the novelty peak was mainly brought by Marshall-1984. The subsequent growth peak after the novelty peak in 1986 occurred in 1988 which was consistent with the top of citation burst of Marshall-1984 in 1988. Meanwhile, the visualization in Figure \ref{fig:peptic_vis} shows the term of \textit{Helicobacter\_Pylori-Associated} emerged in 1988. These observations might explain the growth was partly originated from the novelty brought by Marshall-1984. Another interesting observation is that the growth pattern of publications on Peptic Ulcer from 1986 to 1990 is consistent with the citation growth of Marshall-1984.

\begin{figure}[!ht]
  \centering
  \includegraphics[width=0.9\linewidth]{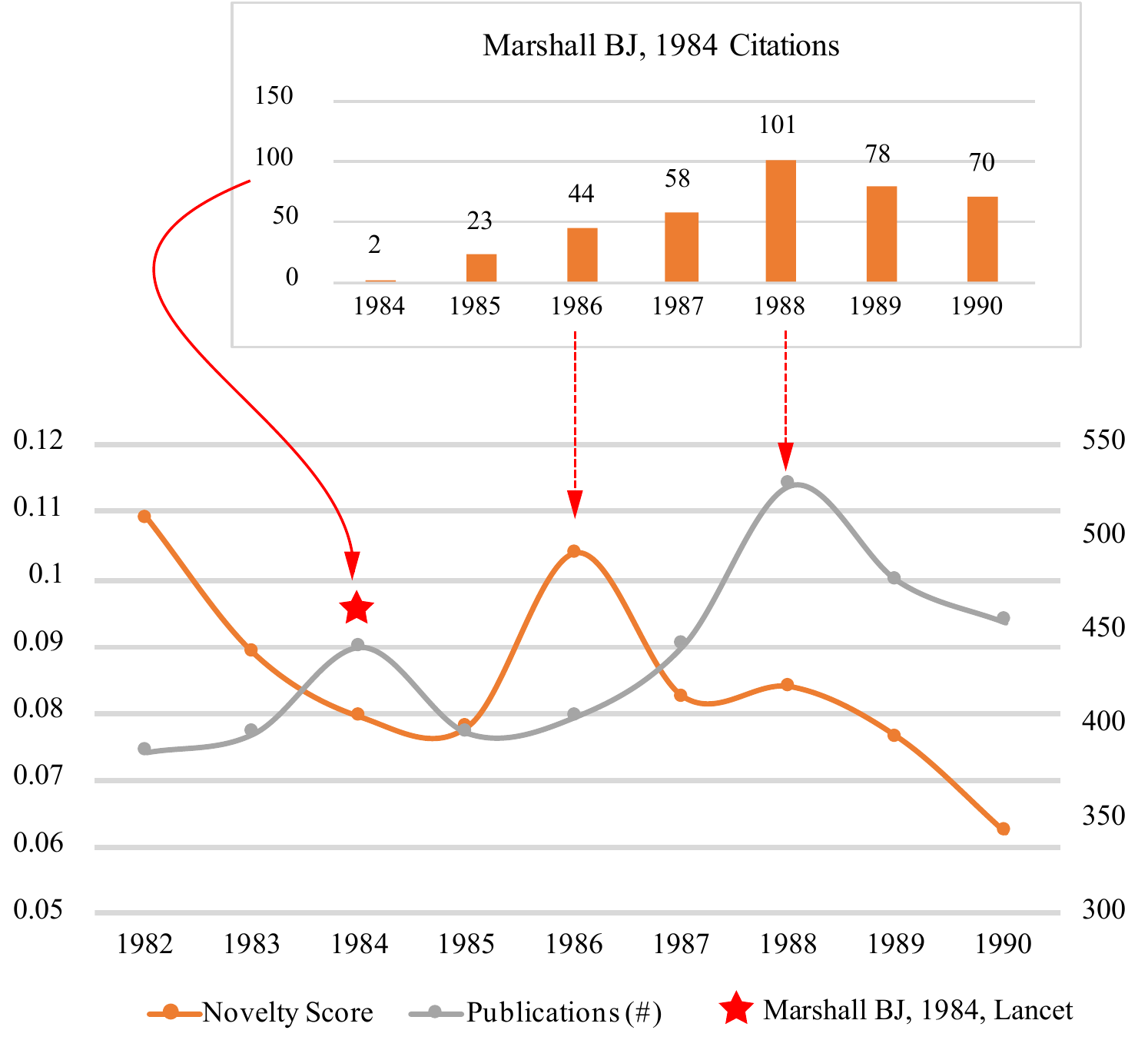}
  \caption{\label{fig:peptic_curve} Co-evolution of the novelty and growth of research on \textit{Peptic Ulcer}.}
\end{figure}


\begin{figure}[!ht]
  \centering
  \includegraphics[width=1\linewidth]{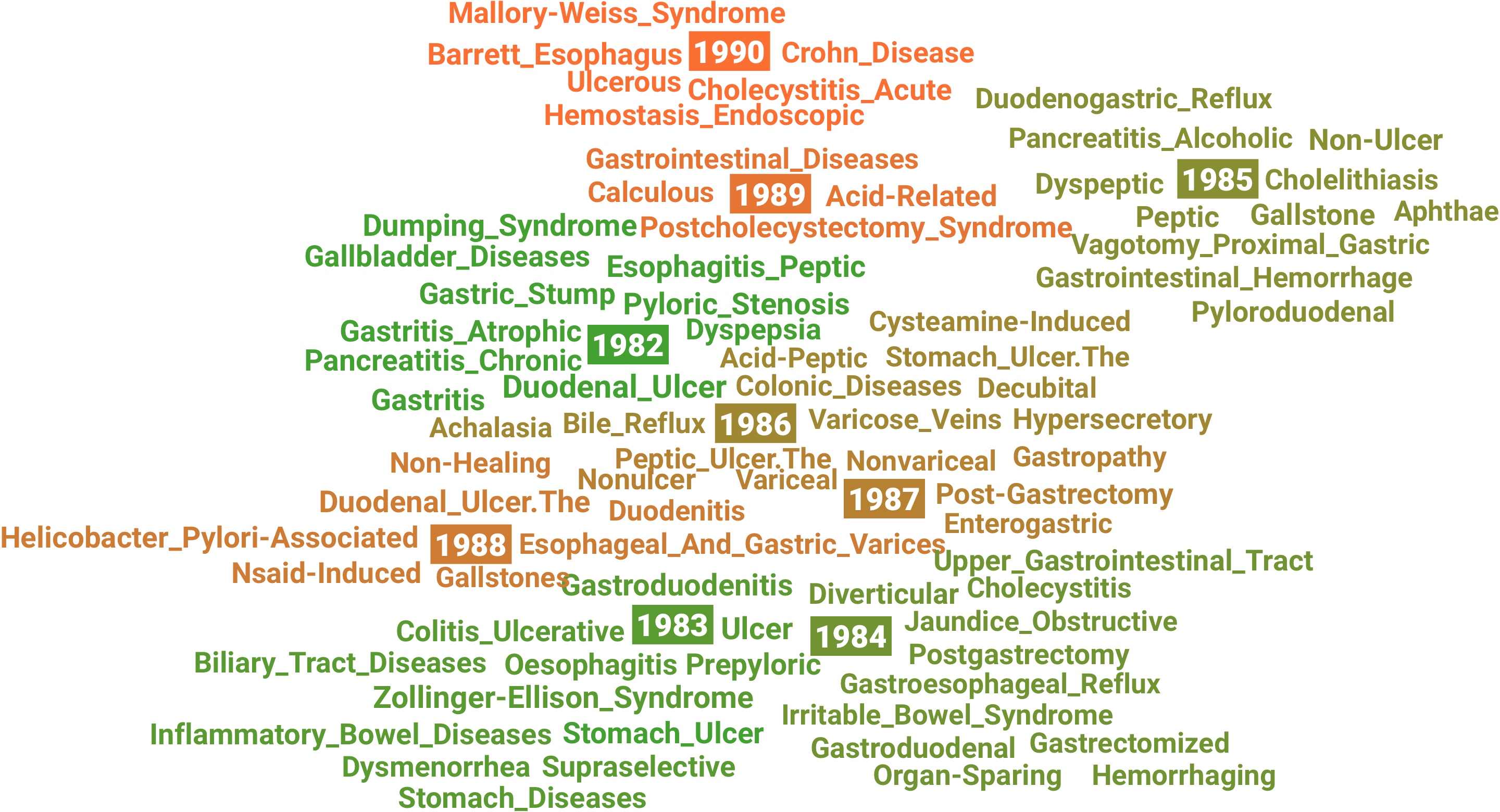}
  \caption{\label{fig:peptic_vis} Visualization of the semantic change in research on \textit{Peptic Ulcer} from 1982 (green) to 1990 (orange).}
\end{figure}
}

\section{Experiment}
We designed a panel data analysis to validate the role of our proposed novelty metric in predicting future growth of research topics and to statistically investigate how it predicts. 


\subsection{Selection of research topics}\label{section:selection}
The focus of this study is to build a novelty metric of research topics and evaluate its predictive effects on scientific growth rather than detecting research topics, so we use MeSH descriptors to characterize research topics {\color{Black}to reduce the complexity associated with implementing the experiment and interpreting the results}. There are two benefits to use MeSH descriptors: (1) MeSH repository is well-maintained by domain experts and is keeping evolving to accommodate dynamics in medical research fields; (2) all the publications in MEDLINE/PubMed are indexed by several MeSH descriptors, so the publication growth on research concepts can be identified by the indexing. Therefore, by retrieving MeSH and MEDLINE/PubMed databases, we may obtain an updating list of research topics and accurate publication growth data regarding the research topics.

However, some of the descriptors have too generic meaning to describe research topics, such as \textit{Animal}\footnote{https://meshb.nlm.nih.gov/record/ui?ui=D000818} and \textit{Child, Preschool}\footnote{https://meshb.nlm.nih.gov/record/ui?ui=D002675}. We need to select the descriptors which have the capacity for describing specific research topics for our experiment. How a descriptor was used by domain experts to index publications may indicate how it can describe a specific research topic. A descriptor may index a publication as either a major topic or a non-major topic. Normally, a scientific publication mainly discusses one specific topic, so if a descriptor was more usually used to index publications as a major topic, the descriptor is more likely to be able to describe a specific topic. Conversely, a descriptor which is more usually to be used as a non-major topic is less likely to be able to describe a specific topic. Therefore, we defined a metric SID (Specific Information of a Descriptor) to quantify a MeSH descriptor’s capacity for describing a specific research topic {\color{Black}which is motivated by the idea of Term Frequency–Inverse Document Frequency (TF-IDF)}, and it is specified as following formula 
\begin{equation}
	\textup{SID}(d_i) = n_{major}(d_i)^{*} \log{\frac{N}{n_{major}(d_i)+n_{non-major}(d_i)}}
\end{equation}
where \(d_i\) is a descriptor, \(n_{major}(d_i)\) is the number of publications in MEDLINE/PubMed which are indexed by descriptor \(d_i\) as major topic, \(n_{non-major}(d_i)\) is the number of publications which are indexed by descriptor \(d_i\) as non-major topic, and \(N\) (=26,759,399) is the total number of publications in MEDLINE/PubMed. {\color{Black}The SID scores of MeSH descriptors ranges between 0 and 259,401 and the average of the SID scores is 2905.} 

{\color{Black}Based on the metric SID and other requirements of the experiment, we decide if a research topic is selected in our experiment by using following criteria:
\begin{enumerate}
    \item The SID score is higher than 1000. {\color{Black}We empirically set the threshold (SID=1000) to remove most of inappropriate cases, even a few appropriate cases may be removed by setting the threshold, but enough appropriate cases would be retained for our statistical investigation by setting the threshold. 10,580 out of 27,804 (38.1\%) MeSH descriptors have a SID score higher than 1000.}
    \item The count of terms describing the research topic in textual training data (i.e. title and abstracts of PubMed) in the observed year is larger than 50. 
    \item It has at least a half of consecutive years within the observed period in which the topic meets the requirement 2. For example, if we conducted a panel data analysis in a period from 1996 to 2005, the research topic should have at least 5 consecutive years in which its terms occurred more than 50 times in textual training data of each year.
\end{enumerate}
}

\subsection{Variables}

\subsubsection{Dependent variable}

The dependent variable is the growth of a research topic within a certain year. In this study, we used a common indicator to characterize the scientific growth of a certain research topic: publication growth \citep{Price1951, ASI:ASI23329}.
The publication growth of a research topic is the growth rate of the number of papers whose topic list has the research topic, i.e., the papers were indexed with the research topic in PubMed.
The growth rate of a certain year is simply calculated as follows:
\begin{equation}
	Growth^{(t)} = \frac{N^{(t)}-N^{(t-1)}}{N^{(t-1)}} * 100
\end{equation} 
where \(N^{(t)}\) is the number of publications within year \(t\).

\subsubsection{Independent variable}

\textit{Novelty}. We measure the novelty of a research topic by our method described in section \ref{section:measure_novelty}. The novelty of a research topic within a certain year is decided by the difference between the semantic properties of the research topic within the year and its historical years. To identify the number of years for retrospecting which makes novelty with strongest ability to predict the growth, we created variables of novelty with 10 different retrospective windows in this experiment, i.e., we created \( \{ novelty^{(t)}(win) : win = 1, 2, ..., 10\} \) for each research topic \( w_i \). We only employ one variable of novelty as an independent variable in each statistical evaluation, and investigate the predictive effects of novelty variables with different retrospective windows.

\textit{Growth}. The growth of a research topic within year $t$ may also be a predictive indicator for the growth within year $t+ \Delta t$ ($\Delta t > 0$).

\subsubsection{Control variables}
The novelty wouldn't be the only factor affecting the growth of science, so our model needs to be adopted to exclude other identifiable factors in our dataset which may affect the growth. Our statistical model included two control variables: research field and age of research topic.

\textit{Research field}. The dynamics of science regarding the number of publications \citep{lietz2013science}
differs from field to field. In this experiment, the research field of a research topic is decided by the main branch of MeSH headings the research topic is under, i.e., the first level of MeSH heading tree \citep{Medcine2017}. There are 16 research fields according to this category list. We define the research topics under multiple categories as inter-field research topics. Thus, we use a set of 17 field dummies (F1--F17) to indicate the research fields.

\textit{Age of research topic}. Another factor that may affect the growth of a research topic is the time it has been established. The growth of a long-established research topic may be more stable and with lower growth rate than a newly established research topic. In this experiment, we use the established date of MeSH headings to indicate the age of corresponding research topics. 

\subsection{Panel data regression model} \label{panel_model}
The data for this experiment is a panel dataset mainly containing growth data and novelty data of the research topic samples spanning from 1991 to 2005. The experiment applied panel data models to investigate relationship between the novelty and growth of research topics. {\color{Black}The fundamental advantage of a panel data set over a cross section is that it will allow us great flexibility in modeling differences in scientific dynamics across individual research topics \citep{Greene2012}}. The basic model for our panel data analysis is a regression model of the form
\begin{equation}
	Growth_{i}^{(t+\Delta)} = \mathbf{z}_i'\boldsymbol{\alpha } + \beta_{1}Novelty_i^{(t)} + \beta_{2}Growth_{i}^{(t)} + \beta_{3}Age_i^{(t)} +\beta_{4}Field_i + \varepsilon_i^{(t)}
\end{equation} 
{\color{Black}where $i = 1,2, ...,N$ is the individual research topic index, $t = 1,2, ...,T$ is the time index. \( \mathbf{z}_i'\boldsymbol{\alpha }\) is the heterogeneity or individual effect of research topics where \(\mathbf{z}_i\) contains a constant term and a set of individual observed variables of research topics, such as field and interdisciplinarity, or unobserved ones, such as specific characteristics;} the dependent variable \( Growth_{i}^{(t+ \Delta t)} \) is the growth of research topic \(w_i\) within a given year \(t + \Delta t: \Delta t=1,...,10\); \(Novelty_i^{(t)}\) is the novelty of the research topic \(w_i\) within year \(t\); \(Field_i\) is the research field of research topic \(w_i\); \(Age_i^{(t)}\) is how many years it has been at \(t\) since the research topic \(w_i\) was established; \(\varepsilon_i^{(t)}\) is the idiosyncratic error.

Research topics are likely to systematically differ from each other and change over time. There are two types of models for capturing cross-sectional heterogeneity: fixed effects model and random effects model \citep{Greene2012}. { \color{Black} Each research topic may have its own individual characteristics that may or may not influence the growth prediction. 

Fixed-effect model controls for time-invariance characteristics that may impact or bias the growth prediction within research topic individuals. To remove the effect of those time-invariant characteristics, intercepts $\mathbf{z}_i'\boldsymbol{\alpha }$ in fixed-effect model are assumed to be different across research topics{\color{Black}, but be constant for each research topic across time}. Thus, the estimated coefficients of fixed-effect model cannot be biased by omitted time-invariant characteristics, but meanwhile, fix-effect model cannot be used to investigate time-invariant causes of the growth of research topics (i.e., $Field$ in our regression model). {\color{Black} The fixed model in our experiment is formulated as
\begin{equation}
	Growth_{i}^{(t+\Delta)} = \alpha_i + \beta_{1}Novelty_i^{(t)} + \beta_{2}Growth_{i}^{(t)} + \beta_{3}Age_i^{(t)} + \varepsilon_i^{(t)}
\end{equation}
where $\alpha_i = \mathbf{z}_i'\boldsymbol{\alpha }$ and $\mathbf{z}_i$ is unobserved. The fixed-effect model takes $\alpha_i$ to be a topic-specific constant term.}

Unlike fixed-effect model, the key rationale behind random-effect model is that the different characteristics across research topics have no influence on their growth predictors.} {\color{Black}In random-effect model, the unobserved but formulated heterogeneity of topic individuals is assumed to be uncorrelated with the independent variables. 
The random-effect model in our experiment is formulated as
\begin{equation}\label{eq:fixed_model}
	Growth_{i}^{(t+\Delta)} = \alpha + u_i + \beta_{1}Novelty_i^{(t)} + \beta_{2}Growth_{i}^{(t)} + \beta_{3}Age_i^{(t)} +\beta_{4}Field_i + \varepsilon_i^{(t)}
\end{equation}
The random-effect model formulates the topic-specific effects as random variables and explores the random heterogeneity of topics in error variables ($u_i$ and $\varepsilon_i^{(t)}$). In Equation \ref{eq:fixed_model}, $u_i$ ($={\mathbf{z}_i'\boldsymbol{\alpha } - E[\mathbf{z}_i'\boldsymbol{\alpha }]}$) is the between-topic error and is constant for topic $i$ through time; $\varepsilon_i^{(t)}$ is within-topic error of topic $i$; $\alpha$ ($=E[\mathbf{z}_i'\boldsymbol{\alpha }]$) is a constant term of the mean of the unobserved heterogeneity.}

{ \color{Black} It is possible that both time-specific and topic-specific effects are not statistically significant. In the case without significant topic-specific effect or time-specific effect, pooled regression model is more efficient. } In pooled regression model, $\mathbf{z}_i$ contains only a constant term and ordinary least squares provides estimates of the common intercept \(\mathbf{z}_i'\boldsymbol{\alpha }\) and slope vector $ \boldsymbol{ \beta }$ ($=(\beta_1, \beta_2, \beta_3, \beta_4)$). {\color{Black}The pooled regression model in our experiment is formulated as 
\begin{equation}
	Growth_{i}^{(t+\Delta)} = \alpha + \beta_{1}Novelty_i^{(t)} + \beta_{2}Growth_{i}^{(t)} + \beta_{3}Age_i^{(t)} +\beta_{4}Field_i + \varepsilon_i^{(t)}
\end{equation}}

These model options bring us to the question of which panel data model to use. It is less likely to decide by general inferences about relationships between variables. Fortunately, several specification tests, including Baltagi test ($F$-test), Breusch-Pagan Lagrange Multiplier test (LM test) and Hausman test, will help us decide which panel data model to use \citep{JSSv027i02, Greene2012}. $F$-test helps us decide between pooled regression model and fixed-effect model. The null hypothesis of $F$-test is that the coefficients for all periods are jointly equal to 0. If the null hypothesis is rejected, time-fixed effects are needed for our panel data, and we should select fixed-effect model than pooled regression model. LM test is used to decide between pooled regression model and random-effect model. LM test's null hypothesis is that variance across entities is 0, which means no significant difference across topics in our data. If the null hypothesis is rejected, we will select random-effect model than pooled regression model. If both fixed and random effects turn out significant, Hausman test is used to decide between fixed-effect model and random-effect model. The null hypothesis of Hausman test is that the preferred model is random effects rather than fixed-effect. We would apply these tests to determine which panel data model is most efficient for our experiment. 

\section{Results}
In this section, we described how we decide the window for measuring novelty which can bring most predictive effects on the future growth of scientific knowledge. Then, we reported a descriptive statistics and correlations of the variables. Lastly, we validated and statistically investigated the predictive effects of novelty on the growth of research topics based on the results of panel data analysis. 

\subsection{Novelty with different windows}

{\color{Black}Before conducting our panel data analysis, we need to identify which window of novelty can produce novelty with strongest predictive effects and the identified novelty metric would be used as independent variable $Novelty^{(t)}$ in our experiment.} As we introduced in Section \ref{section:novelty}, we can set different retrospective windows to measure novelty of research topic within a certain period. Novelty with different windows should be consistent with each other but may have different predictive effects on growth. To identify the window leading to novelty metric with strongest predictive effects, we examined the relations between each pair of $Growth^{(t+\Delta t)}$ and $Novelty^{(t)}(win)$ ($\Delta t=1,2,...,10$; $win=1,2,...,10$) by employing the model described in Section \ref{panel_model}. Thus, we would have 10*10 panel analysis models. Although other one independent variable and two control variables would be involved, we only focus on discussing the independent variable $Novelty$ and the dependent variable $Growth$ in this examination. 

Before discussing and interpreting the output of model estimates, we address some model specification issues. The results of various specification tests ($F$-test, LM test, and Hausman test) introduced in Section \ref{panel_model} suggested {\color{Black}that all of 100 models should accept random-effect model. Therefore, we report results of the random-effect model in Table \ref{table:novelty_windows_random}. The table only provides coefficient estimates, standard errors, significant levels of independent variable $Novelty^{(t)}$ with different windows for dependent variables $Growth^{(t+\Delta t)}$.}

\begin{landscape}
\begin{table}[!htbp]
\footnotesize
\centering
\caption{Estimates the effects of $Novelty^{(t)}$ metics on $Growth^{(t+\Delta t)}$ by random-effect model.}
\label{table:novelty_windows_random}
{ \color{Black}
\begin{tabular}{@{}ccccccccccc@{}}
\toprule
& $G^{(t+1)}$   & $G^{(t+2)}$                                            & $G^{(t+3)}$                                            & $G^{(t+4)}$                                          & $G^{(t+5)}$                                           & $G^{(t+6)}$                                            & $G^{(t+7)}$                                            & $G^{(t+8)}$                                           & $G^{(t+9)}$                                          & $G^{(t+10)}$                                                                                                                                 \\ \midrule
 $Novelty^{(t)}(1)$  & \begin{tabular}[c]{@{}l@{}}0.067 $^{***}$ \\ (0.015)\end{tabular} & \begin{tabular}[c]{@{}l@{}}0.061 $^{***}$ \\ (0.015)\end{tabular} & \begin{tabular}[c]{@{}l@{}}0.061 $^{***}$ \\ (0.014)\end{tabular}  & \begin{tabular}[c]{@{}l@{}}0.080 $^{***}$ \\ (0.014)\end{tabular} & \begin{tabular}[c]{@{}l@{}}0.070 $^{***}$ \\ (0.013)\end{tabular} & \begin{tabular}[c]{@{}l@{}}0.067 $^{***}$ \\ (0.012)\end{tabular} & \begin{tabular}[c]{@{}l@{}}0.073 $^{***}$ \\ (0.012)\end{tabular}  & \begin{tabular}[c]{@{}l@{}}0.078 $^{***}$ \\ (0.012)\end{tabular} & \begin{tabular}[c]{@{}l@{}}0.074 $^{***}$ \\ (0.011)\end{tabular} & \begin{tabular}[c]{@{}l@{}}0.080 $^{***}$ \\ (0.011)\end{tabular} \\
\rule{0pt}{14pt} $Novelty^{(t)}(2)$  & \begin{tabular}[c]{@{}l@{}}0.098 $^{***}$ \\ (0.018)\end{tabular} & \begin{tabular}[c]{@{}l@{}}0.078 $^{***}$ \\ (0.018)\end{tabular} & \begin{tabular}[c]{@{}l@{}}0.083 $^{***}$ \\ (0.017)\end{tabular}  & \begin{tabular}[c]{@{}l@{}}0.091 $^{***}$ \\ (0.016)\end{tabular} & \begin{tabular}[c]{@{}l@{}}0.094 $^{***}$ \\ (0.016)\end{tabular} & \begin{tabular}[c]{@{}l@{}}0.083 $^{***}$ \\ (0.015)\end{tabular} & \begin{tabular}[c]{@{}l@{}}0.093 $^{***}$ \\ (0.014)\end{tabular} & \begin{tabular}[c]{@{}l@{}}0.101 $^{***}$ \\ (0.014)\end{tabular} & \begin{tabular}[c]{@{}l@{}}0.093 $^{***}$ \\ (0.014)\end{tabular} & \begin{tabular}[c]{@{}l@{}}0.096 $^{***}$ \\ (0.013)\end{tabular} \\
\rule{0pt}{14pt} $Novelty^{(t)}(3)$  & \begin{tabular}[c]{@{}l@{}}0.118 $^{***}$ \\ (0.020)\end{tabular} & \begin{tabular}[c]{@{}l@{}}0.084 $^{***}$ \\ (0.019)\end{tabular}  & \begin{tabular}[c]{@{}l@{}}0.098 $^{***}$ \\ (0.019)\end{tabular} & \begin{tabular}[c]{@{}l@{}}0.099 $^{***}$ \\ (0.017)\end{tabular} & \begin{tabular}[c]{@{}l@{}}0.099 $^{***}$ \\ (0.018)\end{tabular} & \begin{tabular}[c]{@{}l@{}}0.096 $^{***}$ \\ (0.016)\end{tabular}  & \begin{tabular}[c]{@{}l@{}}0.102 $^{***}$ \\ (0.016)\end{tabular}  & \begin{tabular}[c]{@{}l@{}}0.108 $^{***}$ \\ (0.015)\end{tabular} & \begin{tabular}[c]{@{}l@{}}0.096 $^{***}$ \\ (0.015)\end{tabular}  & \begin{tabular}[c]{@{}l@{}}0.104 $^{***}$ \\ (0.014)\end{tabular} \\
\rule{0pt}{14pt} $Novelty^{(t)}(4)$  & \begin{tabular}[c]{@{}l@{}}0.117 $^{***}$ \\ (0.021)\end{tabular} & \begin{tabular}[c]{@{}l@{}}0.090 $^{***}$ \\ (0.020)\end{tabular}  & \begin{tabular}[c]{@{}l@{}}0.098 $^{***}$ \\ (0.020)\end{tabular}  & \begin{tabular}[c]{@{}l@{}}0.099 $^{***}$ \\ (0.019)\end{tabular}  & \begin{tabular}[c]{@{}l@{}}0.100 $^{***}$ \\ (0.018)\end{tabular} & \begin{tabular}[c]{@{}l@{}}0.100 $^{***}$ \\ (0.017)\end{tabular}  & \begin{tabular}[c]{@{}l@{}}0.107 $^{***}$ \\ (0.016)\end{tabular}  & \begin{tabular}[c]{@{}l@{}}0.114 $^{***}$ \\ (0.016)\end{tabular} & \begin{tabular}[c]{@{}l@{}}0.100 $^{***}$ \\ (0.016)\end{tabular}  & \begin{tabular}[c]{@{}l@{}}0.112 $^{***}$ \\ (0.015)\end{tabular} \\
\rule{0pt}{14pt} $Novelty^{(t)}(5)$  & \begin{tabular}[c]{@{}l@{}}0.119 $^{***}$ \\ (0.021)\end{tabular} & \begin{tabular}[c]{@{}l@{}}0.097 $^{***}$ \\ (0.021)\end{tabular}  & \begin{tabular}[c]{@{}l@{}}0.101 $^{***}$ \\ (0.020)\end{tabular}  & \begin{tabular}[c]{@{}l@{}}\textbf{0.106 $^{***}$} \\ (0.019)\end{tabular}  & \begin{tabular}[c]{@{}l@{}}0.099 $^{***}$ \\ (0.019)\end{tabular} & \begin{tabular}[c]{@{}l@{}}0.100 $^{***}$ \\ (0.018)\end{tabular}  & \begin{tabular}[c]{@{}l@{}}0.110 $^{***}$ \\ (0.017)\end{tabular}  & \begin{tabular}[c]{@{}l@{}}0.121 $^{***}$ \\ (0.016)\end{tabular} & \begin{tabular}[c]{@{}l@{}}0.102 $^{***}$ \\ (0.016)\end{tabular}  & \begin{tabular}[c]{@{}l@{}}0.114 $^{***}$ \\ (0.015)\end{tabular} \\
\rule{0pt}{14pt} $Novelty^{(t)}(6)$  & \begin{tabular}[c]{@{}l@{}}0.123 $^{***}$ \\ (0.022)\end{tabular} & \begin{tabular}[c]{@{}l@{}}0.095 $^{***}$ \\ (0.021)\end{tabular}  & \begin{tabular}[c]{@{}l@{}}0.106 $^{***}$ \\ (0.021)\end{tabular}  & \begin{tabular}[c]{@{}l@{}}0.102 $^{***}$ \\ (0.020)\end{tabular}  & \begin{tabular}[c]{@{}l@{}}0.102 $^{***}$ \\ (0.019)\end{tabular} & \begin{tabular}[c]{@{}l@{}}0.097 $^{***}$ \\ (0.018)\end{tabular}  & \begin{tabular}[c]{@{}l@{}}0.113 $^{***}$ \\ (0.017)\end{tabular}  & \begin{tabular}[c]{@{}l@{}}0.123 $^{**}$ \\ (0.017)\end{tabular} & \begin{tabular}[c]{@{}l@{}}0.100 $^{***}$ \\ (0.016)\end{tabular}  & \begin{tabular}[c]{@{}l@{}}0.115 $^{***}$ \\ (0.016)\end{tabular} \\
\rule{0pt}{14pt} $Novelty^{(t)}(7)$  & \begin{tabular}[c]{@{}l@{}}0.123 $^{***}$ \\ (0.022)\end{tabular} & \begin{tabular}[c]{@{}l@{}} \textbf{0.109 $^{***}$} \\ (0.022)\end{tabular}  & \begin{tabular}[c]{@{}l@{}}\textbf{0.108 $^{***}$} \\ (0.021)\end{tabular}  & \begin{tabular}[c]{@{}l@{}}0.104 $^{***}$ \\ (0.020)\end{tabular}  & \begin{tabular}[c]{@{}l@{}}\textbf{0.106 $^{***}$} \\ (0.019)\end{tabular} & \begin{tabular}[c]{@{}l@{}}\textbf{0.108 $^{***}$} \\ (0.019)\end{tabular}  & \begin{tabular}[c]{@{}l@{}}0.116 $^{***}$ \\ (0.018)\end{tabular}  & \begin{tabular}[c]{@{}l@{}}\textbf{0.131 $^{***}$} \\ (0.017)\end{tabular} & \begin{tabular}[c]{@{}l@{}}\textbf{0.103 $^{***}$} \\ (0.017)\end{tabular}   & \begin{tabular}[c]{@{}l@{}}\textbf{0.121 $^{***}$} \\ (0.016)\end{tabular} \\
\rule{0pt}{14pt} $Novelty^{(t)}(8)$  & \begin{tabular}[c]{@{}l@{}} \textbf{ 0.126 $^{***}$} \\ (0.022) \end{tabular} & \begin{tabular}[c]{@{}l@{}}0.097 $^{***}$ \\ (0.022)\end{tabular}  & \begin{tabular}[c]{@{}l@{}}0.102 $^{***}$ \\ (0.021)\end{tabular}  & \begin{tabular}[c]{@{}l@{}}0.092 $^{***}$ \\ (0.020)\end{tabular}  & \begin{tabular}[c]{@{}l@{}}0.101 $^{***}$ \\ (0.020)\end{tabular} & \begin{tabular}[c]{@{}l@{}}0.098 $^{***}$ \\ (0.019)\end{tabular}  & \begin{tabular}[c]{@{}l@{}}0.115 $^{***}$ \\ (0.018)\end{tabular}  & \begin{tabular}[c]{@{}l@{}}0.125 $^{***}$ \\ (0.017)\end{tabular} & \begin{tabular}[c]{@{}l@{}}0.101 $^{***}$ \\ (0.017)\end{tabular}   & \begin{tabular}[c]{@{}l@{}}0.118 $^{***}$ \\ (0.016)\end{tabular} \\
\rule{0pt}{14pt} $Novelty^{(t)}(9)$  & \begin{tabular}[c]{@{}l@{}}0.121 $^{***}$ \\ (0.022)\end{tabular} & \begin{tabular}[c]{@{}l@{}}0.089 $^{***}$ \\ (0.022)\end{tabular}  & \begin{tabular}[c]{@{}l@{}}0.101 $^{***}$ \\ (0.021)\end{tabular}  & \begin{tabular}[c]{@{}l@{}}0.092 $^{***}$ \\ (0.020)\end{tabular}  & \begin{tabular}[c]{@{}l@{}}0.100 $^{***}$ \\ (0.019)\end{tabular} & \begin{tabular}[c]{@{}l@{}}0.116 $^{***}$ \\ (0.018)\end{tabular}  & \begin{tabular}[c]{@{}l@{}}\textbf{0.123 $^{***}$} \\ (0.018)\end{tabular}   & \begin{tabular}[c]{@{}l@{}}0.123 $^{***}$ \\ (0.018)\end{tabular}  & \begin{tabular}[c]{@{}l@{}}0.101 $^{***}$ \\ (0.017)\end{tabular}   & \begin{tabular}[c]{@{}l@{}}0.118 $^{***}$ \\ (0.016)\end{tabular} \\
\rule{0pt}{14pt} $Novelty^{(t)}(10)$ & \begin{tabular}[c]{@{}l@{}}0.118 $^{***}$ \\ (0.023)\end{tabular} & \begin{tabular}[c]{@{}l@{}}0.089 $^{***}$ \\ (0.022)\end{tabular}  & \begin{tabular}[c]{@{}l@{}}0.096 $^{***}$ \\ (0.022)\end{tabular}  & \begin{tabular}[c]{@{}l@{}}0.090 $^{***}$ \\ (0.021)\end{tabular}  & \begin{tabular}[c]{@{}l@{}}0.088 $^{***}$ \\ (0.020)\end{tabular} & \begin{tabular}[c]{@{}l@{}}0.100 $^{***}$ \\ (0.019)\end{tabular}   & \begin{tabular}[c]{@{}l@{}}0.112 $^{***}$ \\ (0.018)\end{tabular}   & \begin{tabular}[c]{@{}l@{}}0.123 $^{***}$ \\ (0.018)\end{tabular} & \begin{tabular}[c]{@{}l@{}}0.100 $^{***}$ \\ (0.017)\end{tabular}   & \begin{tabular}[c]{@{}l@{}}0.119 $^{***}$ \\ (0.017)\end{tabular}  \\ \bottomrule

\end{tabular}
}
\begin{tablenotes}
    \item \textit{Note}: $G^{(t+\Delta t)} = Growth^{(t+\Delta t)}$. Standard errors in parentheses.
    \item $^{*}$ $p < .05$; $^{**}$ $p < .01$; $^{***}$ $p < .001$.
\end{tablenotes}

\end{table}
\end{landscape}

{\color{Black}Based on results in Table \ref{table:novelty_windows_random}, the novelty metrics with different windows are consistent as we expected. All of $Novelty^{(t)}(win)$ metrics has positive and significant ($p < 0.001$) coefficients for $Growth^{(t+\Delta t)}$, but the coefficients of $Novelty^{(t)}(win)$ metrics vary in a degree for different $win$s. The highest coefficients of $Novelty^{(t)}(win)$ for each $Growth^{(t+\Delta t)}$ were highlighted in bold. When $win <= 7$, we see an rising trend of coefficients of $Novelty^{(t)}(win)$ metric when $win$ goes larger; When $win>7$, the coefficients of $Novelty^{(t)}(win)$ metrics are relatively stable and slightly decline over time. Therefore, a larger window may not necessarily bring a novelty metric with higher predictive effects on growth, but expanding the window to 7 would provide a novelty metric with higher predictive effects. Based on this analysis, we selected $Novelty^{(t)}(7)$ as novelty metric for further analysis.}

{\color{Black}

\subsection{Descriptive statistics}

Descriptive statistics for the key variables of our panel data are displayed in Table \ref{table:descriptive}. The panel data range from 1996 to 2005. We reported ten sets of panel data for ten dependent variables $Growth^{(t+\Delta t)}$ where $\Delta t = 1,2,...,10$. Based on the selection criteria described in Section \ref{section:selection}, we obtained 7352 topic candidates and 720,512 observations for the ten dependent variables. We removed outlier data records by using z-score ($<3$) of dependent variables $Growth^{(t+\Delta t)}$. After outlier removal, we have 713,912 observations and slightly different numbers of topics for each dependent variable $Growth^{(t+\Delta t)}$ (see Table \ref{table:1996_2005_random_effect} for details). The average growth rate of research topics is 4.63\%, with wide variation as evidenced by the fact that the standard deviation of the growth rate is roughly five times of the mean. Contrarily, research topics have relatively low variation in novelty. 
\begin{landscape}
\begin{table}[!htbp]
\footnotesize
\centering
\caption{Descriptive statistics}
\label{table:descriptive}

{\color{Black}
\begin{tabular}{@{}llllllllllll@{}}
\toprule
  & Variables                 &         $N$          & Mean      & Std. Dev     & Median & Min        & Max      & 1      & 2     & 3     & 4     \\ \midrule
1 & $Growth^{(t+ \Delta t)}$ (\%)     & 713,912     & 4.63      & 22.13       & 2.86   & -99.13     & 271.43 &        &      &       &       \\
2 & $Novelty^{(t)}$ (\%)              & 713,912     & 12.05     & 4.57        & 11.64  & 1.30      & 72.48  & 0.024$^{***}$  &       &       &       \\
3 & $Growth^{(t)}$ (\%)             & 713,912     & 17.43      & 96.79    & 3.65   & -99.13     & 409.38 & 0.032$^{***}$ & 0.004 $^{***}$ &       &       \\
4 & $Age^{(t)} $ (years)             & 713,912     & 23.90     & 13.70        & 30.00  & 1.30     & 72.50  & -0.042$^{***}$ & 0.024$^{***}$    & -0.141$^{***}$ &       \\
5 & $Field$                           & 713,912     & -         &  -          & -      &  -         &  -     & 0.008$^{***}$  & -0.052$^{***}$ & 0.031$^{***}$     & -0.081$^{***}$ \\\bottomrule
\end{tabular}    
}
\begin{tablenotes}
    \item \textit{Note}: $t = 1996, 1992, ..., 2005; \Delta t = 1, 2, ..., 10$.
    \item $^{*}$ $p < .05$; $^{**}$ $p < .01$; $^{***}$ $p < .001$.
\end{tablenotes}
\end{table}
\end{landscape}
$Growth^{(t+\Delta t)}$ is positively correlated with all independent variables ($Novelty^{(t)}$ and $Growth^{(t)}$) and control variables ($Field$ and $Age^{(t)}$), but only the correlation between $Growth^{(t+\Delta t)}$ and $Age^{(t)}$ is negative. $Novelty^{(t)}$ is positively correlated with both $Growth^{(t+\Delta t)}$ and $Growth^{(t)}$, which means the novelty of a research topic in a specific year is correlated with the growth of the research topic in this year as well as its subsequent growth after this year. However, the correlation between $Novelty^{(t)}$ and $Growth^{(t)}$ is very weak compared with $Growth^{(t+\Delta t)}$, which may indicate that more publications don't necessarily bring high novelty, but high novelty may bring rapid publication growth in the future. We further discuss the relationship between $Novelty^{(t)}$ and $Growth^{(t+\Delta t)}$ in the next section.
}

{\color{Black}
\subsection{Panel data regression analysis}

Table \ref{table:1996_2005_random_effect} presents the results of the random-effect models and specification tests, applied to the sample from 1996 to 2005 of our dataset. Each column is the results of a random-effect model whose dependent variables are $Growth^{(t+\Delta t)}$ where $\Delta t=0, 1, ..., 10$ respectively. 

Before interpreting the estimates results, we discuss some model specification issues. The row of `$F$-test' indicates that fixed effects are not needed for all models, i.e., there is no substantial inter-topic variation. The row of `LM test' corresponds to a test of the statistical significance of topic random effects, which can be used to decide whether random effects are needed. The null hypothesis of 0 variances of the topic-specific error is rejected for models of all columns. Accordingly, the random-effect model is preferred for all models. {\color{Black}The reason for why random-effect model is appropriate may be that research topics in our experiment were drawn from a large population (7,311 out of 27,804) and it might be appropriate to model the individual specific constant terms as randomly distributed across research topics.} Thus, we treat random-effect models as our benchmark in this study.
\begin{landscape}
\begin{table}[!htbp]
\footnotesize
\centering

\caption{$Growth^{(t+\Delta t)}$ random-effect models (1996-2005).}
\label{table:1996_2005_random_effect}
{\color{Black}

\begin{tabular}{@{}ccccccccccc@{}} 
\toprule
& $G^{(t+1)}$   & $G^{(t+2)}$                                            & $G^{(t+3)}$                                            & $G^{(t+4)}$                                          & $G^{(t+5)}$                                           & $G^{(t+6)}$                                            & $G^{(t+7)}$                                            & $G^{(t+8)}$                                           & $G^{(t+9)}$                                          & $G^{(t+10)}$
\\ \midrule
\rule{0pt}{16pt} $Novelty^{(t)}$         & \begin{tabular}[c]{@{}l@{}} 0.123$^{***}$ \\(0.022)  \end{tabular} & \begin{tabular}[c]{@{}l@{}} 0.109$^{***}$ \\(0.022)   \end{tabular} & \begin{tabular}[c]{@{}l@{}} 0.108$^{***}$ \\(0.021)   \end{tabular} & \begin{tabular}[c]{@{}l@{}} 0.104$^{***}$ \\(0.020)   \end{tabular} & \begin{tabular}[c]{@{}l@{}} 0.106$^{***}$ \\(0.019)   \end{tabular} & \begin{tabular}[c]{@{}l@{}} 0.108$^{***}$ \\(0.018)   \end{tabular} & \begin{tabular}[c]{@{}l@{}} 0.116$^{***}$ \\(0.018)   \end{tabular} & \begin{tabular}[c]{@{}l@{}} 0.131$^{***}$ \\(0.017)   \end{tabular} & \begin{tabular}[c]{@{}l@{}} 0.103$^{***}$ \\(0.017)   \end{tabular} & \begin{tabular}[c]{@{}l@{}} 0.121$^{***}$ \\(0.016)   \end{tabular}\\
\rule{0pt}{16pt} $Growth^{(t)}$        & \begin{tabular}[c]{@{}l@{}} 0.019$^{***}$ \\(0.001)  \end{tabular} & \begin{tabular}[c]{@{}l@{}} 0.007$^{***}$ \\(0.001)   \end{tabular} & \begin{tabular}[c]{@{}l@{}} 0.004$^{***}$ \\(0.000)   \end{tabular} & \begin{tabular}[c]{@{}l@{}} 0.002$^{***}$ \\(0.000)   \end{tabular} & \begin{tabular}[c]{@{}l@{}} 0.002$^{***}$ \\(0.000)   \end{tabular} & \begin{tabular}[c]{@{}l@{}} 0.002$^{***}$ \\(0.000)   \end{tabular} & \begin{tabular}[c]{@{}l@{}} 0.001$^{*}$ \\(0.000)     \end{tabular} & \begin{tabular}[c]{@{}l@{}} 0.001$^{*}$ \\(0.000)     \end{tabular} & \begin{tabular}[c]{@{}l@{}} 0.001$^{**}$ \\(0.000)    \end{tabular} & \begin{tabular}[c]{@{}l@{}} 0.001$^{*}$ \\(0.000)   \end{tabular}  \\
\rule{0pt}{16pt} $Field$           & \begin{tabular}[c]{@{}l@{}} 1.286$^{***}$ \\(0.259)  \end{tabular} & \begin{tabular}[c]{@{}l@{}} 1.100$^{***}$ \\(0.254)   \end{tabular} & \begin{tabular}[c]{@{}l@{}} 1.355$^{***}$ \\(0.247)   \end{tabular} & \begin{tabular}[c]{@{}l@{}} 0.981$^{***}$ \\(0.237)   \end{tabular} & \begin{tabular}[c]{@{}l@{}} 1.151$^{***}$ \\(0.227)   \end{tabular} & \begin{tabular}[c]{@{}l@{}} 0.688$^{***}$ \\(0.212)   \end{tabular} & \begin{tabular}[c]{@{}l@{}} 0.622$^{**}$ \\(0.203)    \end{tabular} & \begin{tabular}[c]{@{}l@{}} 0.528$^{**}$ \\(0.197)    \end{tabular} & \begin{tabular}[c]{@{}l@{}} 0.348 \\(0.190)     \end{tabular} & \begin{tabular}[c]{@{}l@{}} -0.038 \\(0.180)    \end{tabular} \\
\rule{0pt}{16pt} $Age^{(t)}$             & \begin{tabular}[c]{@{}l@{}} -0.107$^{***}$ \\(0.007) \end{tabular} & \begin{tabular}[c]{@{}l@{}} -0.102$^{***}$ \\(0.006) \end{tabular} & \begin{tabular}[c]{@{}l@{}} -0.084$^{***}$ \\(0.006) \end{tabular} & \begin{tabular}[c]{@{}l@{}} -0.080$^{***}$ \\(0.006)  \end{tabular} & \begin{tabular}[c]{@{}l@{}} -0.079$^{***}$ \\(0.006) \end{tabular} & \begin{tabular}[c]{@{}l@{}} -0.056$^{***}$ \\(0.005) \end{tabular} & \begin{tabular}[c]{@{}l@{}} -0.049$^{***}$ \\(0.005) \end{tabular} & \begin{tabular}[c]{@{}l@{}} -0.029$^{***}$ \\(0.005) \end{tabular} & \begin{tabular}[c]{@{}l@{}} -0.026$^{***}$ \\(0.005) \end{tabular} & \begin{tabular}[c]{@{}l@{}} -0.036$^{***}$ \\(0.005) \end{tabular}\\
\rule{0pt}{10pt} $F$-Value          & 310.02         & 117.91           & 75.12            & 60.03           & 65.05           & 42.10            & 32.82           & 23.47           & 19.59            & 29.42            \\
\rule{0pt}{10pt} $R^2$              & 0.017            & 0.007             & 0.004             & 0.003             & 0.004             & 0.002             & 0.003             & 0.001             & 0.001             & 0.003             \\
\rule{0pt}{10pt} Adjusted $R^2$     & 0.017            & 0.007             & 0.004             & 0.003            & 0.004            & 0.002             & 0.002             & 0.001            & 0.001             & 0.002             \\
\rule{0pt}{10pt} $F$-test           & 0.956             & 0.978              & 0.972              & 0.968              & 0.911              & 0.844              & 0.805              & 0.787              & 0.781              & 0.777              \\
\rule{0pt}{10pt} LM test & 47.19 $^{***}$        & 31.67 $^{***}$         & 27.15 $^{***}$         & 41.34 $^{***}$         & 68.01 $^{***}$         & 133.49 $^{***}$        & 188.63 $^{***}$        & 198.64 $^{***}$        & 238.82 $^{***}$        & 410.87 $^{***}$        \\
\rule{0pt}{10pt} Hausman test       & 407.05 $^{***}$       & 544.48 $^{***}$        & 497.79 $^{***}$        & 588.26 $^{***}$        & 360.44 $^{***}$        & 228.14 $^{***}$        & 220.30 $^{***}$        & 126.81 $^{***}$        & 246.79 $^{***}$        & 799.53 $^{***}$\\
\rule{0pt}{10pt} $N$ Obs. & 71,554 &  71,553 & 71,543 & 71,530 & 71,469 & 71,408 &  71,323 & 71,192 & 71,323 & 71,017 \\
\rule{0pt}{10pt} $N$ Topics & 7,311 & 7,313 & 7,314 & 7,316 & 7,313 & 7,313 & 7,313 & 7,313 & 7,313 & 7,313 
        \\ \bottomrule
\end{tabular}

}
\begin{tablenotes}
    \item \textit{Note}: $Novelty^{(t)} = Novelty^{(t)}(7)$; $G^{(t+\Delta t)} = Growth^{(t+\Delta t)}$; Standard errors in parentheses.
    \item $^{*}$ $p < .05$; $^{**}$ $p < .01$; $^{***}$ $p < .001$.
\end{tablenotes}
\end{table}
\end{landscape}

Variables of $Novelty^{(t)}$ and $Age^{(t)}$ are consistent across all models for their significance levels ($p<0.001$). The significance levels of $Growth^{(t)}$ and $Field$ decline as $\Delta t$ becomes larger when $\Delta t > 6$. Only $Age$ has negative effects on future growth. For coefficients, $Novelty^{(t)}$ is only variable that has consistent coefficients across all models and doesn't show an obvious changing pattern. Except $Novelty^{(t)}$, results concerning other independent and control variables don't have consistent coefficients across models, but their changing patterns differ from each other. The coefficients of $Growth^{(t)}$ drops greatly when $\Delta t>1$, so the growth of a research topic may only have predictive effects on the subsequent growth in the fairly near future. The effect estimate of two control variables $Field$ and $Age^{(t)}$ decline gradually as $\Delta t$ becomes larger. These two variables characterize some individual characteristics of research topics, so the characteristics of a research topic may not have lasting effects on its growth. Regarding significance levels and coefficient estimates, $Novelty^{(t)}$ has relatively stable and long-term predictive effects on future growth in science. 

{\color{Black}The relatively low $R_2$ and adjusted $R_2$ results in Table \ref{table:1996_2005_random_effect} indicate that much variation of scientific growth haven't been explained by our model. This may be partly because of the complexity of the temporal pattern of novelty and growth. Based on the results shown in Table \ref{table:1996_2005_random_effect}, $Novelty^{(t)}$ has predictive effects on each year at a population level. However, $Novelty^{(t)}$ may only accelerate growth in certain year(s) rather than every year in the future based our observation in case studies, so the predictive effects of $Novelty^{(t)}$ may spread out over years in the future. The spread may be roughly even according to the consistent coefficients of $Novelty^{(t)}$. Additionally, $R_2$ results have a decreasing trend over time, which may be mainly caused by other variables except $Novelty^{(t)}$ according to the decrease of their significance levels and coefficients over time.}

For the purpose of checking robustness and differences to subsample and time-periods, we consider three alternative time-periods for investigation: 1991 to 1995, 1996 to 2000, and 2001 to 2005. The results for these three time-periods are reported in Table \ref{table:1991_1995_random_effect}, Table \ref{table:1996_2000_random_effect} and Table \ref{table:2001_2005_random_effect} respectively. These tables shows slightly different effect estimates of $Novelty^{(t)}$ on $Growth^{(t+\Delta t)}$. In the time-period from 1991 to 1995, $Novelty^{(t)}$ has positive and significant effect estimates in most of models ($\Delta t=3,4,...,10$) with consistent coefficients, but it has no significant effect on $Growth^{(t+1)}$ and has low coefficient on $Growth^{(t+2)}$ with relatively low significance level ($p<0.05)$). In the time-period from 1996 to 2000, $Novelty^{(t)}$ has positive and significant effect estimates in all models, but the coefficients are not consistent across models which shows a declining pattern. In the time-period from 2001 to 2005, $Novelty^{(t)}$ show consistent effect estimates only when $\Delta t>4$. It has relatively low coefficients on $Growth^{(t+2)}$ and $Growth^{(t+3)}$ with relatively low significance level ($p<0.05$) and has no significant effects on $Growth^{(t+1)}$ and $Growth^{(t+4)}$. $Novelty^{(t)}$ has similar predictive effects on $Growth^{(t+\Delta t)}$ only when $\Delta t > 4$ across these three time-periods.


\begin{landscape}
\begin{table}[!htbp]
\footnotesize
\centering

\caption{$Growth^{(t+\Delta t)}$ random-effect models (1991-1995).}
\label{table:1991_1995_random_effect}
{ \color{Black}

\begin{tabular}{@{}ccccccccccc@{}} 
\toprule
& $G^{(t+1)}$   & $G^{(t+2)}$                                            & $G^{(t+3)}$                                            & $G^{(t+4)}$                                          & $G^{(t+5)}$                                           & $G^{(t+6)}$                                            & $G^{(t+7)}$                                            & $G^{(t+8)}$                                           & $G^{(t+9)}$                                          & $G^{(t+10)}$
\\ \midrule
\rule{0pt}{16pt} $Novelty^{(t)}$         & \begin{tabular}[c]{@{}l@{}} 0.025 \\(0.037)  \end{tabular} & \begin{tabular}[c]{@{}l@{}} 0.069$^{*}$ \\(0.035)   \end{tabular} & \begin{tabular}[c]{@{}l@{}} 0.167$^{***}$ \\(0.033)   \end{tabular} & \begin{tabular}[c]{@{}l@{}} 0.215$^{***}$ \\(0.032)   \end{tabular} & \begin{tabular}[c]{@{}l@{}} 0.237$^{***}$ \\(0.032)   \end{tabular} & \begin{tabular}[c]{@{}l@{}} 0.261$^{***}$ \\(0.034)   \end{tabular} & \begin{tabular}[c]{@{}l@{}} 0.271$^{***}$ \\(0.033)   \end{tabular} & \begin{tabular}[c]{@{}l@{}} 0.227$^{***}$ \\(0.033)   \end{tabular} & \begin{tabular}[c]{@{}l@{}} 0.229$^{***}$ \\(0.032)   \end{tabular} & \begin{tabular}[c]{@{}l@{}} 0.207$^{***}$ \\(0.031)   \end{tabular}\\
\rule{0pt}{16pt} $Growth^{(t)}$        & \begin{tabular}[c]{@{}l@{}} 0.021$^{***}$ \\(0.001)  \end{tabular} & \begin{tabular}[c]{@{}l@{}} 0.004$^{***}$ \\(0.000)   \end{tabular} & \begin{tabular}[c]{@{}l@{}} 0.002$^{***}$ \\(0.000)   \end{tabular} & \begin{tabular}[c]{@{}l@{}} 0.001$^{**}$ \\(0.000)   \end{tabular} & \begin{tabular}[c]{@{}l@{}} 0.001 \\(0.000)   \end{tabular} & \begin{tabular}[c]{@{}l@{}} 0.001$^{*}$ \\(0.000)   \end{tabular} & \begin{tabular}[c]{@{}l@{}} 0.001 \\(0.000)     \end{tabular} & \begin{tabular}[c]{@{}l@{}} 0.000 \\(0.000)     \end{tabular} & \begin{tabular}[c]{@{}l@{}} 0.001$^{*}$ \\(0.000)    \end{tabular} & \begin{tabular}[c]{@{}l@{}} 0.000 \\(0.000)   \end{tabular}  \\
\rule{0pt}{16pt} $Field$           & \begin{tabular}[c]{@{}l@{}} 1.419$^{***}$ \\(0.407)  \end{tabular} & \begin{tabular}[c]{@{}l@{}} 1.426$^{***}$ \\(0.384)   \end{tabular} & \begin{tabular}[c]{@{}l@{}} 1.448$^{***}$ \\(0.362)   \end{tabular} & \begin{tabular}[c]{@{}l@{}} 1.657$^{***}$ \\(0.350)   \end{tabular} & \begin{tabular}[c]{@{}l@{}} 1.629$^{***}$ \\(0.346)   \end{tabular} & \begin{tabular}[c]{@{}l@{}} 1.946$^{***}$ \\(0.370)   \end{tabular} & \begin{tabular}[c]{@{}l@{}} 1.292$^{***}$ \\(0.365)    \end{tabular} & \begin{tabular}[c]{@{}l@{}} 1.613$^{***}$ \\(0.364)    \end{tabular} & \begin{tabular}[c]{@{}l@{}} 1.652$^{***}$ \\(0.344)     \end{tabular} & \begin{tabular}[c]{@{}l@{}} 1.787$^{***}$ \\(0.345)    \end{tabular} \\
\rule{0pt}{16pt} $Age^{(t)}$             & \begin{tabular}[c]{@{}l@{}} -0.203$^{***}$ \\(0.011) \end{tabular} & \begin{tabular}[c]{@{}l@{}} -0.171$^{***}$ \\(0.011) \end{tabular} & \begin{tabular}[c]{@{}l@{}} -0.135$^{***}$ \\(0.010) \end{tabular} & \begin{tabular}[c]{@{}l@{}} -0.092$^{***}$ \\(0.010)  \end{tabular} & \begin{tabular}[c]{@{}l@{}} -0.043$^{***}$ \\(0.010) \end{tabular} & \begin{tabular}[c]{@{}l@{}} -0.046$^{***}$ \\(0.010) \end{tabular} & \begin{tabular}[c]{@{}l@{}} -0.050$^{***}$ \\(0.010) \end{tabular} & \begin{tabular}[c]{@{}l@{}} -0.061$^{***}$ \\(0.010) \end{tabular} & \begin{tabular}[c]{@{}l@{}} -0.057$^{***}$ \\(0.010) \end{tabular} & \begin{tabular}[c]{@{}l@{}} -0.082$^{***}$ \\(0.010) \end{tabular}\\
\rule{0pt}{10pt} $F$-Value    & 423.55         & 93.38     & 63.74     & 43.33    &  23.62      & 27.82   & 26.43           & 24.81           & 29.39        & 35.74            \\
\rule{0pt}{10pt} $R^2$              & 0.052            & 0.012             & 0.008            & 0.006             & 0.003             & 0.004             & 0.003             & 0.003             & 0.004             & 0.005             \\
\rule{0pt}{10pt} Adjusted $R^2$     & 0.052            & 0.012             & 0.008             & 0.005            & 0.003            & 0.004             & 0.003             & 0.003            & 0.004             & 0.004             \\
\rule{0pt}{10pt} $F$-test           & 0.866             & 0.841              & 0.774              & 0.736              & 0.729              & 0.807              & 0.784              & 0.781              & 0.723              & 0.787              \\
\rule{0pt}{10pt} LM test & 155.58 $^{***}$        & 204.15 $^{***}$         & 246.88$^{***}$         & 273.43 $^{***}$         & 303.80 $^{***}$         & 156.45 $^{***}$        & 172.80 $^{***}$        & 175.19 $^{***}$        & 256.90 $^{***}$        & 159.69 $^{***}$        \\
\rule{0pt}{10pt} Hausman test       & 50.19 $^{***}$       & 303.36 $^{***}$        & 132.85 $^{***}$        & 40.60 $^{***}$        & 105.81 $^{***}$        & 78.73 $^{***}$        & 30.93 $^{***}$        & 37.92 $^{***}$        & 18.14 $^{***}$        & 2.36 \\
\rule{0pt}{10pt} $N$ Obs. & 31,041 & 31,036 & 31,026 & 31,025 & 31,011 & 31,026 &  31,012 & 30,963 & 30856 & 30,713 \\
\rule{0pt}{10pt} $N$ Topics & 6,301 & 6,302 & 6,303 & 6,303 & 6,302 & 6,303 & 6,303 & 6,303 & 6,302 & 6,305
        \\ \bottomrule
\end{tabular}
}
\begin{tablenotes}
    \item \textit{Note}: $Novelty^{(t)} = Novelty^{(t)}(7)$; $G^{(t+\Delta t)} = Growth^{(t+\Delta t)}$; Standard errors in parentheses.
    \item $^{*}$ $p < .05$; $^{**}$ $p < .01$; $^{***}$ $p < .001$.
\end{tablenotes}
\end{table}
\end{landscape}
\begin{landscape}
\begin{table}[!htbp]
\footnotesize
\centering

\caption{$Growth^{(t+\Delta t)}$ random-effect models (1996-2000).}
\label{table:1996_2000_random_effect}
{ \color{Black}

\begin{tabular}{@{}ccccccccccc@{}} 
\toprule
& $G^{(t+1)}$   & $G^{(t+2)}$                                            & $G^{(t+3)}$                                            & $G^{(t+4)}$                                          & $G^{(t+5)}$                                           & $G^{(t+6)}$                                            & $G^{(t+7)}$                                            & $G^{(t+8)}$                                           & $G^{(t+9)}$                                          & $G^{(t+10)}$
\\ \midrule
\rule{0pt}{16pt} $Novelty^{(t)}$         
&     \begin{tabular}[c]{@{}l@{}} 0.227$^{***}$ \\(0.034)  \end{tabular} & \begin{tabular}[c]{@{}l@{}} 0.180$^{***}$ \\(0.033)   \end{tabular} & \begin{tabular}[c]{@{}l@{}} 0.167$^{***}$ \\(0.032)   \end{tabular} & \begin{tabular}[c]{@{}l@{}} 0.176$^{***}$ \\(0.030)   \end{tabular} & \begin{tabular}[c]{@{}l@{}} 0.126$^{***}$ \\(0.030)   \end{tabular} & \begin{tabular}[c]{@{}l@{}} 0.145$^{***}$ \\(0.027)   \end{tabular} & \begin{tabular}[c]{@{}l@{}} 0.142$^{***}$ \\(0.026)   \end{tabular} & \begin{tabular}[c]{@{}l@{}} 0.139$^{***}$ \\(0.025)   \end{tabular} & \begin{tabular}[c]{@{}l@{}} 0.123$^{***}$ \\(0.024)   \end{tabular} & \begin{tabular}[c]{@{}l@{}} 0.110$^{***}$ \\(0.023)   \end{tabular}\\
\rule{0pt}{16pt} $Growth^{(t)}$        
&     \begin{tabular}[c]{@{}l@{}} 0.016$^{***}$ \\(0.001)  \end{tabular} & \begin{tabular}[c]{@{}l@{}} 0.005$^{***}$ \\(0.001)   \end{tabular} & \begin{tabular}[c]{@{}l@{}} 0.003$^{***}$ \\(0.001)   \end{tabular} & \begin{tabular}[c]{@{}l@{}} 0.002$^{**}$ \\(0.001)   \end{tabular} & \begin{tabular}[c]{@{}l@{}} 0.001 \\(0.001)   \end{tabular} & 
    \begin{tabular}[c]{@{}l@{}} 0.002$^{**}$ \\(0.001)   \end{tabular} &
    \begin{tabular}[c]{@{}l@{}} 0.000 \\(0.001)     \end{tabular} & 
    \begin{tabular}[c]{@{}l@{}} 0.001 \\(0.001)     \end{tabular} & 
    \begin{tabular}[c]{@{}l@{}} 0.001 \\(0.001)    \end{tabular} & 
    \begin{tabular}[c]{@{}l@{}} 0.001 \\(0.001)   \end{tabular}  \\
\rule{0pt}{16pt} $Field$           
& \begin{tabular}[c]{@{}l@{}} 1.710 $^{***}$ \\(0.392)  \end{tabular} & \begin{tabular}[c]{@{}l@{}} 1.104$^{**}$ \\(0.382)   \end{tabular} & \begin{tabular}[c]{@{}l@{}} 1.370$^{***}$ \\(0.366)   \end{tabular} & \begin{tabular}[c]{@{}l@{}} 1.448$^{***}$ \\(0.343)   \end{tabular} & \begin{tabular}[c]{@{}l@{}} 1.762$^{***}$ \\(0.340)   \end{tabular} & \begin{tabular}[c]{@{}l@{}} 1.011$^{**}$ \\(0.312)   \end{tabular} & \begin{tabular}[c]{@{}l@{}} 1.103$^{***}$ \\(0.297)    \end{tabular} & \begin{tabular}[c]{@{}l@{}} 1.316$^{***}$ \\(0.282)    \end{tabular} & \begin{tabular}[c]{@{}l@{}} 0.614$^{*}$ \\(0.272)     \end{tabular} & \begin{tabular}[c]{@{}l@{}} 0.437 \\(0.254)    \end{tabular} \\
\rule{0pt}{16pt} $Age^{(t)}$             
& \begin{tabular}[c]{@{}l@{}} -0.118$^{***}$ \\(0.010) \end{tabular} & \begin{tabular}[c]{@{}l@{}} -0.112$^{***}$ \\(0.010) \end{tabular} & \begin{tabular}[c]{@{}l@{}} -0.097$^{***}$ \\(0.010) \end{tabular} & \begin{tabular}[c]{@{}l@{}} -0.089$^{***}$ \\(0.009)  \end{tabular} & \begin{tabular}[c]{@{}l@{}} -0.106$^{***}$ \\(0.009) \end{tabular} & \begin{tabular}[c]{@{}l@{}} -0.068$^{***}$ \\(0.008) \end{tabular} & \begin{tabular}[c]{@{}l@{}} -0.053$^{***}$ \\(0.008) \end{tabular} & \begin{tabular}[c]{@{}l@{}} -0.021$^{**}$ \\(0.007) \end{tabular} & \begin{tabular}[c]{@{}l@{}} -0.016$^{*}$ \\(0.007) \end{tabular} & \begin{tabular}[c]{@{}l@{}} -0.015$^{*}$ \\(0.007) \end{tabular}\\
\rule{0pt}{10pt} $F$-Value    & 145.97         &  62.38     & 41.75     & 39.93    &  49.05     & 29.61   & 22.16           & 14.87         & 9.22       & 8.12            \\
\rule{0pt}{10pt} $R^2$              & 0.017            & 0.007             & 0.005            & 0.005             & 0.006             & 0.003             & 0.003            & 0.002             & 0.001             & 0.001             \\
\rule{0pt}{10pt} Adjusted $R^2$     & 0.017            & 0.007             & 0.005            & 0.005             & 0.006             & 0.003             & 0.003            & 0.002             & 0.001             & 0.001             \\
\rule{0pt}{10pt} $F$-test    
    & 0.930  
    & 0.905        
    & 0.842
    & 0.774
    & 0.813
    & 0.765
    & 0.750
    & 0.727
    & 0.696
    & 0.612              \\
\rule{0pt}{10pt} LM test 
    & 88.68 $^{***}$        
    & 98.93 $^{***}$         
    & 132.72$^{***}$         
    & 220.88 $^{***}$         
    & 148.78 $^{***}$         
    & 212.49 $^{***}$        
    & 251.50 $^{***}$        
    & 273.38 $^{***}$        
    & 349.18 $^{***}$        
    & 541.73 $^{***}$        \\
\rule{0pt}{10pt} Hausman test       
    & 298.66 $^{***}$        
    & 229.95 $^{***}$         
    & 133.56$^{***}$         
    & 84.75 $^{***}$         
    & 15.15 $^{***}$         
    & 11.07 $^{***}$        
    & 104.91 $^{***}$        
    & 63.66 $^{***}$        
    & 97.05 $^{***}$        
    & 32.34 $^{***}$ \\
\rule{0pt}{10pt} $N$ Obs. 
& 34,296 
& 34,295 
& 34,288 
& 34,277 
& 34,243 
& 34,206 
& 34,177 
& 34,133 
& 34,079 
& 33,991 \\
\rule{0pt}{10pt} $N$ Topics 
& 6,952 
& 6,925 
& 6,924 
& 6,923 
& 6,923 
& 6,924 
& 6,924 
& 6,923 
& 6,923 
& 6,924
        \\ \bottomrule
\end{tabular}
}
\begin{tablenotes}
    \item \textit{Note}: $Novelty^{(t)} = Novelty^{(t)}(7)$; $G^{(t+\Delta t)} = Growth^{(t+\Delta t)}$; Standard errors in parentheses.
    \item $^{*}$ $p < .05$; $^{**}$ $p < .01$; $^{***}$ $p < .001$.
\end{tablenotes}
\end{table}
\end{landscape}

\begin{landscape}
\begin{table}[!htbp]
\footnotesize
\centering

\caption{$Growth^{(t+\Delta t)}$ random-effect models (2001-2005).}
\label{table:2001_2005_random_effect}
{ \color{Black}

\begin{tabular}{@{}ccccccccccc@{}} 
\toprule
& $G^{(t+1)}$   & $G^{(t+2)}$                                            & $G^{(t+3)}$                                            & $G^{(t+4)}$                                          & $G^{(t+5)}$                                           & $G^{(t+6)}$                                            & $G^{(t+7)}$                                            & $G^{(t+8)}$                                           & $G^{(t+9)}$                                          & $G^{(t+10)}$
\\ \midrule
\rule{0pt}{16pt} $Novelty^{(t)}$         
&     \begin{tabular}[c]{@{}l@{}} 0.040 \\(0.028)  \end{tabular} 
& \begin{tabular}[c]{@{}l@{}} 0.058$^{*}$ \\(0.027)   \end{tabular} & \begin{tabular}[c]{@{}l@{}} 0.053$^{*}$ \\(0.025)   \end{tabular} & \begin{tabular}[c]{@{}l@{}} 0.036 \\(0.024)   \end{tabular} & \begin{tabular}[c]{@{}l@{}} 0.072$^{**}$ \\(0.022)   \end{tabular} & \begin{tabular}[c]{@{}l@{}} 0.068$^{**}$ \\(0.022)   \end{tabular} & \begin{tabular}[c]{@{}l@{}} 0.079$^{***}$ \\(0.022)   \end{tabular} & \begin{tabular}[c]{@{}l@{}} 0.107$^{***}$ \\(0.022)   \end{tabular} & \begin{tabular}[c]{@{}l@{}} 0.078$^{***}$ \\(0.021)   \end{tabular} & \begin{tabular}[c]{@{}l@{}} 0.106$^{***}$ \\(0.021)   \end{tabular}\\
\rule{0pt}{16pt} $Growth^{(t)}$        
&     \begin{tabular}[c]{@{}l@{}} 0.031$^{***}$ \\(0.001)  \end{tabular} & \begin{tabular}[c]{@{}l@{}} 0.009$^{***}$ \\(0.001)   \end{tabular} & \begin{tabular}[c]{@{}l@{}} 0.006$^{***}$ \\(0.001)   \end{tabular} & \begin{tabular}[c]{@{}l@{}} 0.005$^{***}$ \\(0.001)   \end{tabular} & \begin{tabular}[c]{@{}l@{}} 0.003$^{***}$ \\(0.001)   \end{tabular} & 
    \begin{tabular}[c]{@{}l@{}} 0.003$^{***}$ \\(0.001)   \end{tabular} &
    \begin{tabular}[c]{@{}l@{}} 0.002$^{***}$ \\(0.001)     \end{tabular} & 
    \begin{tabular}[c]{@{}l@{}} 0.002$^{***}$ \\(0.001)     \end{tabular} & 
    \begin{tabular}[c]{@{}l@{}} 0.002$^{***}$ \\(0.001)    \end{tabular} & 
    \begin{tabular}[c]{@{}l@{}} 0.002$^{**}$ \\(0.001)   \end{tabular}  \\
\rule{0pt}{16pt} $Field$           
& \begin{tabular}[c]{@{}l@{}} 0.891 $^{**}$ \\(0.392)  \end{tabular} & \begin{tabular}[c]{@{}l@{}} 1.160$^{***}$ \\(0.382)   \end{tabular} & \begin{tabular}[c]{@{}l@{}} 1.433$^{***}$ \\(0.366)   \end{tabular} & \begin{tabular}[c]{@{}l@{}} 0.646$^{*}$ \\(0.340)   \end{tabular} & \begin{tabular}[c]{@{}l@{}} 0.513$^{*}$ \\(0.312)   \end{tabular} & \begin{tabular}[c]{@{}l@{}} 0.289 \\(0.297)    \end{tabular} & \begin{tabular}[c]{@{}l@{}} 0.139 \\(0.282)    \end{tabular} & \begin{tabular}[c]{@{}l@{}} -0.220 \\(0.272)     \end{tabular} &
    \begin{tabular}[c]{@{}l@{}} 0.097 \\(0.272)     \end{tabular} &
    \begin{tabular}[c]{@{}l@{}} -0.396 \\(0.254)    \end{tabular} \\
\rule{0pt}{16pt} $Age^{(t)}$             
& \begin{tabular}[c]{@{}l@{}} -0.101$^{***}$ \\(0.008) \end{tabular} & \begin{tabular}[c]{@{}l@{}} -0.104$^{***}$ \\(0.008) \end{tabular} & \begin{tabular}[c]{@{}l@{}} -0.064$^{***}$ \\(0.007) \end{tabular} & \begin{tabular}[c]{@{}l@{}} -0.044$^{***}$ \\(0.007)  \end{tabular} & \begin{tabular}[c]{@{}l@{}} -0.037$^{***}$ \\(0.006) \end{tabular} & \begin{tabular}[c]{@{}l@{}} -0.035$^{***}$ \\(0.006) \end{tabular} & \begin{tabular}[c]{@{}l@{}} -0.033$^{***}$ \\(0.006) \end{tabular} & \begin{tabular}[c]{@{}l@{}} -0.032$^{**}$ \\(0.006) \end{tabular} & \begin{tabular}[c]{@{}l@{}} -0.032$^{*}$ \\(0.006) \end{tabular} & \begin{tabular}[c]{@{}l@{}} -0.038$^{*}$ \\(0.006) \end{tabular}\\
\rule{0pt}{10pt} $F$-Value    
    & 349.73  
    & 104.62        
    & 53.69
    & 27.94
    & 21.16
    & 18.09
    & 15.72
    & 16.68
    & 16.02
    & 19.93
            \\
\rule{0pt}{10pt} $R^2$             
    & 0.037  
    & 0.011        
    & 0.006
    & 0.003
    & 0.002
    & 0.002
    & 0.002
    & 0.002
    & 0.002
    & 0.002
             \\
\rule{0pt}{10pt} Adjusted $R^2$     
& 0.037  
    & 0.011        
    & 0.006
    & 0.003
    & 0.002
    & 0.002
    & 0.002
    & 0.002
    & 0.002
    & 0.002
            \\
\rule{0pt}{10pt} $F$-test    
    & 0.807  
    & 0.875        
    & 0.818
    & 0.769
    & 0.658
    & 0.637
    & 0.620
    & 0.629
    & 0.652
    & 0.759              \\
\rule{0pt}{10pt} LM test 
    & 245.23 $^{***}$        
    & 196.11 $^{***}$         
    & 212.50$^{***}$         
    & 278.46 $^{***}$         
    & 485.59 $^{***}$         
    & 551.15 $^{***}$        
    & 571.51 $^{***}$        
    & 522.49 $^{***}$        
    & 538.41 $^{***}$        
    & 587.29 $^{***}$        \\
\rule{0pt}{10pt} Hausman test       
    & 343.66 $^{***}$        
    & 858.42 $^{***}$         
    & 514.12$^{***}$         
    & 303.46 $^{***}$         
    & 96.31$^{***}$         
    & 57.01 $^{***}$        
    & 56.52 $^{***}$        
    & 27.54 $^{***}$        
    & 245.29 $^{***}$        
    & 1065.34 $^{***}$ \\
\rule{0pt}{10pt} $N$ Obs. 
& 36,695 
& 36,687 
& 36,673
& 36,647
& 36,644 
& 36,630 
& 36,610
& 36,563 
& 36,491 
& 36,388 \\
\rule{0pt}{10pt} $N$ Topics 
& 7,407 
& 7,408 
& 7,408 
& 7,406 
& 7,404 
& 7,404 
& 7,402 
& 7,399 
& 7,398 
& 7,400
        \\ \bottomrule
\end{tabular}
}
\begin{tablenotes}
    \item \textit{Note}: $Novelty^{(t)} = Novelty^{(t)}(7)$; $G^{(t+\Delta t)} = Growth^{(t+\Delta t)}$; Standard errors in parentheses.
    \item $^{*}$ $p < .05$; $^{**}$ $p < .01$; $^{***}$ $p < .001$.
\end{tablenotes}
\end{table}
\end{landscape}
}
{\color{Black}

\section{Discussion}

We validated that our proposed novelty metric has predictive effects on publication growth in science. Whereas novelty has predictive effects on far-future growth with high consistency across different time-periods, we see its different predictive effects on near-future growth across different time-periods. This may be because knowledge diffusion in scientific communities takes time \citep{Bettencourt2008}. Although sometimes new knowledge may obtain attention and acceptance of scientific communities once it got published, the impact of the new knowledge manifested by new publications still takes time which is necessary for going through a scientific publication cycle \citep{bjork2005lifecycle}. In our case study on \textit{Peptic Ulcer}, the first major article \citep{MARSHALL19841311} for the paradigm-shifting discovery of \textit{Helicobacter pylori} had been published in 1984, but it created enough impact in science for improving the novelty degree of research on \textit{Peptic Ulcer} until 1986 and the rapid growth of publications followed by the novelty occurred until 1988. It is reasonable to see a time lag between the occurrence of novelty and the occurrence of rapid growth due to the novelty, which may be caused by the differences in knowledge diffusion or length of publication cycles across research topics. However, how to explain the different time lags for different time-periods at the population level still needs further study.

In our three case studies, we observed some individual characteristics of research topics which affect the role of novelty in scientific development. Both \textit{Ebola} in Case I and \textit{H1N1} in Case II are viruses once caused global disease outbreaks. Because of the notable social factor, we observed that the novelty burst and rapid growth of publications occurred in the same year of their outbreaks (2014 for \textit{Ebola} and 2009 for \textit{H1N1}). However, we cannot confirm the interactive mechanisms between these variables in this study. The infectious disease outbreaks might accelerate the progress of research with high novelty and then the produced novel knowledge might boost the growth immediately, or they might boost the novelty and growth simultaneously. The research topic \textit{Peptic Ulcer} in Case III may not have the perceived urgency to the public as much as the other two cases, but it has an established research consensus in the research community. A research community with a firm consensus naturally tends to be resistant to radical novelty. However, the resistance to novel ideas can be helpful for the emergence of the novelty of research topics. As we described in Section \ref{section:case_peptic}, the novelty brought by the discovery of \textit{Helicobacter pylori} emerged in 1986, even the idea of the discovery hadn't been accepted by the scientific community. Even though the resistance might not hinder the emergence of novelty, it might slow down the growth brought the novelty. Only a small citation burst (1988) of Marshall-1984 could be observed \citep{CHEN2009191} before the discovery had been accepted by peptic the ulcer research community in the 1990s \citep{pincock2005nobel}. 

The temporal patterns of growth and novelty in science are profoundly complex. At first, the time lags between novelty and the growth may vary across research topics and time-periods, which we have seen in our case studies (individual-level) and panel data analysis (population-level). The reasons may vary widely, such as social factors, the speed of knowledge diffusion, firmness of established paradigm, and so on. Secondly, a growth burst may be related to multiple novelty peaks, and a novelty peak may cause multiple growth bursts. For example, we only show one growth burst brought by the novelty of the discovery of \textit{Helicobacter pylori}, but actually the novelty has accelerated multiple growth bursts after 1990.

Another interesting point worth mentioning is the effects of the \textit{law of conformity} on the relationship between novelty and growth. The basic idea of the law of conformity is frequently used words change at slower rates, which has been found in non-scientific dataset \citep{hamilton_diachronic_2016}. The law may weaken the predictive effects of the novelty metric, especially for research topics with a large amount publications. However, in a recent study \citep{YAN201876}, they found no evidence to support the \textit{law of conformity} in biomedical literature. Our correlation analysis on $Novelty^{(t)}$ and $Growth^{(t)}$ is in line with this conclusion. Although they are significantly correlated, it is a positive and weak correlation (0.004). It can also be explained by Kuhn's theory on \textit{Scientific Revolutions} \citep{Kuhn:1970}: although normal science is rigid and scientific communities are close-knit, fundamental novelties of fact and theory bring about paradigm change.  

Finally, we believe our investigative study on novelty metric has valuable practical implications on understanding and predicting scientific evolution, but we also think the novelty metric is a simplified way to represent and learn the semantic changes of research topics. This simplification may extract most salient feature from the semantic changes instead of full features. Although we statistically validated the theory of the predictive effects of novelty on growth in science, the relatively low $R^2$ values in our panel data analysis results indicate that much variation hasn't been explained by our panel data model. Other unexplored features that can be derived from semantic changes of research topics should provide more valuable early signs of scientific dynamics, which leave plenty of opportunities for future studies.
}

\section{Conclusions}

In this study, we have proposed a new method to quantify the novelty of a research topic within a certain period through temporal embedding learning. By applying the method of measuring novelty in a large dataset, we have found statistical evidence of the predictive effects of novelty on rapid growth in science. A research topic with highly novel content manifested by scientific literature is likely to subsequently gain growth in publication. {\color{Black}Also, we found different temporal patterns of relationship between novelty and growth from both the individual and population level.}

The proposed novelty metric can computationally represent the new knowledge created by scientific community by utilizing large-scale text data of scientific literature. The statistical investigation of this study shows that the metric is promising in predicting scientific knowledge growth and emerging trends in science. The temporal embedding learning method has encouraging potential in assisting scientists, science policy makers and the public in understanding scientific dynamics. The representation learning methods and investigative results should be helpful for further development of prediction model by artificial intelligence techniques.

This study focuses on proposing a metric for quantifying novelty in science and investigating the relations between novelty and growth in science. There are many new challenges and opportunities beyond the scope of this study. For example, the novelty metric proposed in this study is used to quantify `generic' novelty without distinguishing different types of novelty, such as novelty generated by scientific revolution with limited prior development or by putting an existing research to a new use/context. We validated the predictive effects of the proposed novelty metric and it is promising in predicting emerging topics, but we cannot employ this metric alone to nominate emerging topics. How common are the predictive effects of novelty on growth overall beyond medical areas? What are the other attributes of scientific knowledge growth, how to quantify these attributes, and how do they interact with the attribute of novelty? There are many other potentially valuable techniques that we may incorporate for various applications, including network science, topic modeling, and deep learning. 

We conclude that novelty, as an essential characteristic of science development, has the potential to reveal the underlying mechanisms of the growth of scientific knowledge. Additionally, our proposed novelty metric could represent the novelty in science and has a role in predicting future growth of science.

\section*{Conflict of Interest Statement}

We appreciate The authors declare that the research was conducted in the absence of any commercial or financial relationships that could be construed as a potential conflict of interest.

\section*{Author Contributions}

JH designed the study and conducted the experiment and wrote the first version of the manuscript;
CC improved the result interpretation and the methods of the study and revised the manuscript.

\section*{Funding}
The work is supported by the National Science Foundation (Award Number: 1633286).

\section*{Acknowledgments}
We appreciate all the constructive comments and suggestions from both reviewers. We also thank Xiumei Li and Yang Li at Lebow College of Business, Drexel University for their help in conducting panel data analysis in this study.  

\bibliographystyle{frontiersinSCNS_ENG_HUMS} 


\end{document}